\documentclass[11pt]{article}

\usepackage{latexsym}
\usepackage{psfig}
\usepackage{url}

\newcommand{\Sup}{\mathit{Sup}}
\newcommand{\CC}{\mathit{CC}}
\newcommand{\dff}{\mathit{Def}}
\newcommand{\mod}{\mathit{Mod}}
\newcommand{\pth}{\mathit{path}}
\newcommand{\dom}{\mathit{Dom}}
\newcommand{\Tr}{\mathbf{t}}
\newcommand{\Fa}{\mathbf{f}}

\newcommand{\datan}{\mathrm{DATALOG}^{\neg}}
\newcommand{\data}{\mathrm{data}}

\newcommand{\cc}{\mathit{cc}}

\newcommand{\size}{\mathit{size}}

\newcommand{\indx}{\mathit{index}}
\newcommand{\core}{\mathit{core}}
\newcommand{\cl}{\mathit{cl}}

\newcommand{\HU}{\mathit{HU}}
\newcommand{\pr}{\mathit{Pr}}
\newcommand{\gr}{\mathit{gr}}

\newcommand{\at}{\mathit{At}}
\newcommand{\plc}{\mathit{PS}}
\newcommand{\eplc}{\mathit{PS}+}
\newcommand{\HB}{\mathit{HB}}
\newcommand{\plcp}{\mathit{PS}+}

\newcommand{\col}{\mathit{color}}
\newcommand{\gcl}{\mathit{gcl}}
\newcommand{\tc}{\mathit{tc}}
\newcommand{\vc}{\mathit{vc}}

\newcommand{\nq}{\mathit{nq}}
\newcommand{\invc}{\mathit{invc}}
\newcommand{\hc}{\mathit{hc}}
\newcommand{\hcp}{\mathit{hc\_perm}}
\newcommand{\hce}{\mathit{hc\_edge}}
\newcommand{\start}{\mathit{start}}
\newcommand{\visit}{\mathit{visit}}
\newcommand{\clr}{\mathit{clrd}}

\newcommand{\vtx}{\mathit{vtx}}
\newcommand{\edge}{\mathit{edge}}
\newcommand{\und}{{\_}}

\newcommand{\rra}{\rightarrow}
\newcommand{\Ra}{\Rightarrow}
\newcommand{\La}{\Leftarrow}
\newcommand{\lla}{\leftarrow}

\newtheorem{theorem}{Theorem}[section]

\newtheorem{corollary}[theorem]{Corollary}

\newtheorem{proposition}[theorem]{Proposition}

\begin{document}
\ \\
\begin{center}
{\large\bf Propositional satisfiability in declarative 
programming\footnote{Parts of this paper appeared appeared 
in Proceedings of AAAI-2000 \cite{et00a} and in Proceedings of 
KI-2001 \cite{et01a}.}}
\ \\
\ \\
Deborah East\\
{\small Department of Computer Science\\
Southwest Texas State University\\
San Marcos, TX 78666, USA}\\
\ \\
and\\
\ \\
Miros\l aw Truszczy\'nski\\
{\small Department of Computer Science\\
University of Kentucky\\
Lexington, KY 40506-0046, USA}\\
\end{center}

\begin{abstract}
Answer-set programming (ASP) paradigm is a way of using logic to solve 
search problems. Given a search  problem, to solve it one designs a 
theory in the logic so that {\em models} of this theory represent 
problem solutions. To compute a solution to a problem one needs to
compute a model of the corresponding theory. Several answer-set 
programming formalisms have been developed on the basis of logic 
programming with the semantics of stable models. In this paper we show 
that also the logic of predicate calculus gives rise to effective 
implementations of the ASP paradigm, similar in spirit to logic 
programming with stable model semantics and with a similar scope of 
applicability. Specifically, we propose two logics based on predicate 
calculus as formalisms for encoding search problems. We show that 
the expressive power of these logics is given by the class NP-search. 
We demonstrate how to use them in programming and develop computational
tools for model finding. In the case of one of the logics our techniques 
reduce the problem to that of propositional satisfiability and allow 
one to use off-the-shelf satisfiability solvers. The language of the 
other logic has more complex syntax and provides explicit means to model 
some high-level constraints. For theories in this logic, we designed 
our own solver that takes advantage of the expanded syntax. We present 
experimental results demonstrating computational effectiveness of the 
overall approach.
\end{abstract}

\section{Introduction}

Logic is most commonly used in declarative programming and {\em 
computational} knowledge representation as follows. To solve a problem, 
we represent its general constraints and relevant background knowledge 
as a theory. We express a specific instance of the problem as a 
formula. We then use proof techniques to decide whether this formula 
follows from the theory. A proof of the formula or, more precisely, 
variable substitutions used by the proof, determine a solution which, 
in most cases is represented by a ground term. This use of logic in 
programming and computing stems from the pioneering work by Robinson 
\cite{ro65}, Green \cite{gr69a} and Kowalski \cite{ko74}. It led to 
the establishment of {\em logic programming} as, arguably, the most 
prominent and most broadly accepted logic-based declarative 
programming formalism, and to the development of Prolog as its 
implementation by Colmerauer and his group \cite{ckpr}. 

Recently, researchers proposed an alternative way in which logic 
can be used in computation \cite{mt99,nie99}. {\em Answer-set 
programming} (ASP) is an approach to declarative programming in which 
one represents a computational problem as a theory in some logic so 
that {\em models} of this theory, and not proofs or variable
substitutions, 
represent problem solutions. In ASP, finding models rather than 
proofs is a primary computational task and serves as a foundation 
for a uniform processing mechanism. 

ASP was first explicitly identified as a declarative modeling and 
computational paradigm in knowledge representation
applications of {\em normal} logic programs (programs with negation 
in the bodies) \cite{mt99,nie99}. That work suggested that in order 
to model a problem, a programmer should write a normal logic program 
so that some distinguished {\em models} of the program 
represent solutions. The most broadly accepted 2-valued semantics of 
normal logic programs is the semantics of {\em stable} models 
\cite{gl88}. Thus, under that proposal, a normal logic program {\em solves} a 
problem if answers to the problem are represented by {\em stable} 
models of the program and can be recovered from them. We refer to 
this variant of logic programming as {\em stable logic programming} 
(SLP). It is clear that SLP is a specific instance of the
ASP paradigm described earlier.

In general, stable models are infinite and, unless one can devise for 
them some finitary representation schema, they cannot be computed. 
To overcome this difficulty, it is common to restrict attention in SLP
to finite DATALOG$^\neg$ programs, that is, finite programs 
without function symbols. A stable model of a logic program is, by 
definition, an Herbrand model. Thus, if a program is a finite 
DATALOG$^\neg$ program, its stable models are finite. Moreover, there 
are algorithms to compute stable models of finite DATALOG$^\neg$ 
programs, as well as fast implementations of these algorithms, including 
{\em smodels} \cite{ns00}, {\em dlv} \cite{elmps98}\footnote{In fact, 
{\em dlv} implements an algorithm to compute stable models of 
disjunctive logic programs, a more general task.}, {\em cmodels} 
\cite{cmodels} and {\em assat} \cite{lz02}. These implementations compute 
stable models of DATALOG$^\neg$ programs in two steps. First, an input 
program is {\em grounded}, that is, replaced by a program consisting of 
ground clauses only (hence, it is a propositional program) that has the 
same stable models as the original one. Second, stable models of the 
{\em ground} program are computed by means of search algorithms 
similar to the Davis-Putnam algorithm for satisfiability (SAT) testing
\cite{ns00,elmps98} or, after some additional preprocessing, by SAT 
solvers \cite{cmodels,lz02}.

SLP is a declarative programming formalism particularly well suited for 
representing and solving search problems in the class NP-search, as it 
provides a {\em uniform} solution to each search problem in that class 
\cite{mr00}. Specifically, for every 
search problem $\Pi$ in the class NP-search there is a finite 
DATALOG$^\neg$ program $P_\Pi$, and an encoding schema that represents 
every instance $I$ of $\Pi$ as a finite collection $\data(I)$ of ground 
atoms from the Herbrand base of $P_\Pi$, such that stable models of 
the program $P_\Pi \cup \data(I)$ specify all solutions to the instance 
$I$ of the problem $\Pi$. Problems such as scheduling, planning, 
diagnosis, abductive reasoning, product configuration, versions of
bounded model checking, as well as a broad spectrum of combinatorial 
problems, are members of the class NP-search and so, admit a uniform 
solution in SLP. A recent research monograph 
\cite{baral03} presents SLP solutions for several of these problems, 
especially those that appear in knowledge representation. The 
combination of uniform encoding, high expressive power 
and fast algorithms for computing stable models makes SLP an attractive 
and effective declarative programming formalism. Extensions of the 
language of DATALOG$^\neg$ with explicit representations for constraints 
involving cardinalities and other aggregates, a corresponding 
generalization of the notion of a stable model, and modifications in 
algorithms to compute (generalized) stable models
resulted in even more effective programming 
and computing systems \cite{ns00,sns02}.

While the notion of the answer-set programming paradigm has first 
explicitly appeared in the context of SLP,
it is clear that the way in which propositional 
SAT solvers are used fits well the ASP paradigm. For 
instance, in the satisfiability planning \cite{kms96}, problems are 
encoded as propositional theories so that models determine valid plans. 
SAT solvers are then used to compute them. Other similar uses of SAT
solvers abound. Our goal in this paper is to extend this general idea 
and show that predicate logic with the Herbrand-model semantics, together 
with SAT solvers and their extensions as processing engines, gives rise 
to effective implementations of the ASP paradigm similar in spirit to 
SLP and with a similar scope of applicability. A specific logic we 
propose to this end is a modification of the logic of propositional 
schemata \cite{kms96}. The key concept is that of a {\em 
data-program} pair $(D,P)$ to represent separately a search problem 
$\Pi$ by the {\em program} component $P$ of $(D,P)$, and concrete 
instances of $\Pi$ by the {\em data} part $D$ of $(D,P)$. To define 
the semantics of data-program pairs, we restrict the class 
of Herbrand models of the theory $D\cup P$ to those Herbrand models that 
satisfy a version of Reiter's Closed-World Assumption. We refer to our 
logic as the logic of propositional schemata with Closed-World 
Assumption or, simply, as the logic of propositional schemata. We 
denote this logic by $\plc$.

The logic $\plc$ offers only basic logical connectives to help model
problem constraints. We extend logic $\plc$ to support direct
representation of constraints involving cardinalities. Examples of
such constraints are: "at least $k$ elements from the list must be
in the model" or "exactly $k$ elements from the list must be in the
model". They appear commonly in statements of constraint satisfaction
problems. We also extend the language of the logic $\plc$ with 
Horn rules and use them as means to compute consequences of collections of
ground facts (in particular, to compute transitive closures of binary
relations). We refer to this new logic as {\em extended logic of
propositional schemata} (with Closed-World Assumption) and denote it 
by $\plcp$.

In the paper we study basic properties of the logic $\plc$ and 
observe that they extend to the logic $\plcp$, as well. We show 
that the logic $\plc$ is nonmonotonic, identify sources of
nonmonotonicity and its implications. We demonstrate the use of 
the logic $\plc$ as a representation language be developing programs 
for several search problems. We characterize the class of problems 
that can be solved by programs in the logic $\plc$. To this end, 
we define a formal setting for the study of the expressive power of 
ASP formalisms. We establish that the expressive power of the logic 
$\plc$ is equal to the class NP-search. In particular, it is the same 
as the expressive power of SLP.

As we pointed out, in the logic $\plc$, to solve a problem for a 
particular instance we represent the problem and the instance by 
a data-program pair so that Herbrand models of the data-program
pair correspond to problem solutions. Consequently, the basic 
computational task is that of computing Herbrand models. It can
be accomplished in a similar two-step process to that used in
computing stable models. Given a finite data-program pair, we first
ground it (compute its equivalent propositional representation) and
then find models of the ground theory obtained. 

For grounding, we implemented a program, {\em psgrnd} that, given a 
data-program pair produces an equivalent (with the same models) 
propositional theory. If the input data-program pair is in the language
of the basic logic $\plc$, one of the options of {\em psgrnd} creates 
a theory in the DIMACS format and allows one to use for solving
``off-the-shelf'' SAT solvers. In this way, our logic $\plc$ and our
program {\em psgrnd} provide a programming front-end for SAT solvers
greatly facilitating their use.

If a data-program pair contains higher-level constructs proper to
the logic $\plcp$, one still can use propositional solvers for
processing by first compiling cardinality and closure constraints
to propositional logic and then using SAT solvers to compute models.
However, propositional representations of constraints involving 
cardinalities or closure operators are usually very large and the sizes 
of the compiled theories limit the effectiveness of satisfiability 
checkers, even the most advanced ones, as processing engines. Thus, we 
argue for an alternative approach to design solvers
specifically tailored to the syntax of the logic $\plcp$. To this end, 
we propose a ``target'' propositional logic for the logic $\plcp$. In 
this logic, cardinality and closure
constraints have explicit representations and, therefore, do not need 
to be compiled any further. We develop a satisfiability solver, {\em
aspps}, for the propositional logic $\plcp$ and use it as the processing 
back-end for the logic $\plcp$ and {\em psgrnd}. Our solver is designed 
along the same lines as most satisfiability solvers implementing the 
Davis-Putnam algorithm but it takes a direct advantage of the 
cardinality and closure constraints explicitly present in the language. 

Experimental results on the performance of the overall approach are 
highly encouraging. On one hand, we demonstrate the ease and 
effectiveness of using off-the-shelf SAT solvers to attack search 
problems. On the other hand, we show that significant gains in 
performance can be obtained by developing more general solvers, such 
as {\em aspps}, capable of directly processing some classes of more
complex constraints than those that can be expressed as clauses.
In particular, our solver {\em aspps} is competitive with current
SLP solvers such as {\em smodels} and with complete SAT solvers such
as {\em zchaff} and {\em satz}. In fact, in several instances we 
considered, it was faster. Our work demonstrates that building 
propositional solvers capable of processing high-level constraints 
is a promising research direction for the area of propositional 
satisfiability.

Several interrelated factors motivate us in this work. First, we want
to provide an effective programming front-end that would capitalize
on dramatic improvements in the performance of satisfiability solvers
and would facilitate their use as computational tools. In recent years,
researchers have developed several fast implementations of the basic
Davis-Putnam method such as {\em satz} \cite{la97}, {\em relsat}
\cite{bs97} and, most recently, {\em chaff} \cite{mmz01,chaff}. A
renewed interest in local-search techniques resulted in highly
effective (albeit incomplete) satisfiability checkers such as {\em
WALKSAT} \cite{skc94}, capable of handling large CNF theories,
consisting of millions of clauses. These advances make computing
Herbrand models of predicate-logic theories feasible. Second, the
use of propositional semantics also makes it easy to expand the
basic language with constructs to explicitly represent high-level
constraints and to exploit ideas developed in the area of
propositional satisfiability to design algorithms for computing
models of ground theories in expanded languages. Third, the
semantics of Herbrand models of predicate theories, being essentially
the semantics of propositional logic, is broadly known and much less
complex than the semantics of stable models of logic programs.
Consequently, the task of programming may in many cases be simpler
and the overall approach may gain broader acceptance.

\section{Basic logic $\plc$}
\label{plc}

In this section, we introduce the logic $\plc$ that provides a 
theoretical basis for a declarative programming front-end to 
satisfiability solvers and facilitates their use.  Syntactically, the 
logic $\plc$ is a fragment of first-order logic without function symbols
(or, in other words, predicate logic). Specifically, the language of the 
logic $\plc$ consists of:
\begin{enumerate}
\item infinite denumerable sets $R$, $C$ and $V$ of {\em relation}, 
{\em constant} and {\em variable} symbols
\item symbols $\bot$ and $\top$ (later interpreted always as falsity 
and truth)
\item boolean connectives $\wedge$, $\vee$ and $\rra$, the
universal and existential quantifiers, and  punctuation symbols 
`(', `)' and `,'.
\end{enumerate}
In the paper, following the example of logic programming, we adopt 
the convention that upper-case letters denote variables and 
lower-case letters stand for constants. 

Constant and variable symbols are the only {\em terms} of the
language. Constants are the only {\em ground} terms and they form 
the {\em Herbrand universe} of the language. {\em Atoms} are 
expressions of the form $p(t_1,\ldots,t_n)$, where $p$ is an $n$-ary 
relation symbol from $R$ and $t_i$, $1\leq i\leq n$, are terms. An 
atom $p(t_1,\ldots, t_n)$ is {\em ground} if all its terms are ground. 
The set of all ground atoms forms the {\em Herbrand base} of the 
language.

In the logic $\plc$, we restrict the use of existential quantifiers. 
Let us consider a tuple of terms $(t_1,\ldots,t_n)$ and let $X_1,
\ldots, X_k$ be pairwise distinct variables such that each $X_i$, 
$1\leq i\leq k$, appears in the tuple $(t_1, \ldots,t_n)$ exactly
once. An expression of the form 
\[
\exists X_1,\ldots,X_k\hspace{2mm} p(t_1,\ldots,t_n)
\]
is an {\em e-atom}. For instance, the expression 
$\exists X, Z\hspace{2mm} p(X,Y,Z,c)$ is an e-atom while the expression
$\exists X, Z\hspace{2mm} p(X,X,Z,c)$ is not. In the logic $\plc$, 
existential quantifiers appear {\em exclusively} in e-atoms. 

The requirement that each $X_i$, $1\leq i\leq k$, appears in the 
tuple $(t_1,\ldots, t_n)$ exactly once is not essential and can be 
lifted. We adopt it as it allows us to simplify the notation for
e-atoms. Namely, we write an e-atom
\[
\exists X_1,\ldots,X_k\hspace{2mm} p(t_1,\ldots,t_n)
\]
as
\[
p(t_1',\ldots,t'_n),
\]
where $t'_i=t_i$, if $t_i$ is not one of the variables
$X_1,\ldots,X_k$, and $t_i=\mbox{`$\und$'}$ (underscore), otherwise.
For instance, we write an e-atom $\exists X, Z\hspace{2mm} p(X,Y,Z,c)$ as
$p(\und,Y,\und,c)$.

The {\em only} formulas we allow in the logic $\plc$ are {\em rules},
that is, formulas
\[
\forall X_1,\ldots, X_k (A_1\wedge \ldots\wedge A_m \rra
B_1\vee\ldots\vee B_n),
\]
where all $A_i$, $1\leq i\leq m$, and $B_j$, $1\leq j\leq n$, are
atoms, none of $A_i$'s is an e-atom (in other words, e-atoms do 
not appear in the antecedents of clauses) and $X_1,\ldots, X_k$ are
the free variables appearing in  $A_1\ldots,A_m$ and
$B_1,\ldots,B_n$. If $m=0$, we replace the conjunct in the antecedent
of the clause with the symbol $\top$. If $n=0$, we replace the empty
disjunct in the consequent of the clause with the symbol $\bot$.

As usual, we drop the universal quantifiers from the rule
notation. For instance, to denote the rule 
\[
\forall X,Y,Z (p(X,Y)\wedge p(Y,Z) \rra q(\und,X)\vee q(Y,\und)\vee
r(Z)),
\]
which is already a shorthand for 
\[
\forall X,Y,Z (p(X,Y)\wedge p(Y,Z) \rra \exists W\hspace{2mm} q(W,X)\vee 
\exists W\hspace{2mm} q(Y,W)\vee r(Z)),
\]
we write
\[
p(X,Y)\wedge p(Y,Z) \rra q(\und,X)\vee q(Y,\und)\vee r(Z).
\]
This notation is reminiscent of that commonly used for {\em clauses} in
predicate logic. 
There is a key difference, though. Some of the atoms in the consequent
of a rule may be e-atoms (as it is the case in the example just given). 
Thus, unlike in the case of clauses, a rule of the logic $\plc$ may 
contain the existential quantifier in the consequent.

The last syntactic notion we need is that of a theory. A {\em theory} 
in the logic $\plc$ (or, a $\plc$ theory) is any {\em finite} collection 
of rules that contains at least one occurrence of a constant symbol.

To recap the discussion of the syntax of the logic $\plc$, it is 
essentially a fragment of the syntax the fist-order logic with the 
following restrictions and caveats: (1) function symbols are not
allowed, (2) rules are the only formulas, (3) theories are finite and 
contain at least one constant symbol, and (4) through the use of 
notational conventions, the quantifiers are only implicitly present in 
the language.

The difference between the logic $\plc$ and the corresponding fragment 
of the first-order logic is in the way we interpret theories. Namely, 
we view a theory $T$ as a representation of a {\em certain class of 
models of $T$} and not as a representation of logical consequences of 
$T$. In fact, due to the way we use the logic $\plc$, the concept of 
provability plays virtually no role in logic $\plc$. This is an essential 
departure from the classical first-order logic perspective. 

Specifically, we assign to a $\plc$ theory $T$ a collection of its 
{\em Herbrand models}. The concepts of an Herbrand interpretation, of 
truth in an interpretation and of an Herbrand model, that we use in 
the paper, are standard (for details, we refer to any text in logic, 
for instance, \cite{ns92}). Here we will only introduce some 
necessary notation.
Let $T$ be $\plc$ theory. We denote by $\HU(T)$ the {\em Herbrand 
universe} of $T$, that is, in our case, the set of all constants that 
{\em appear} in $T$. By the definition of a $\plc$ theory, this set is
not empty and finite. We denote by $\HB(T)$ the {\em Herbrand base} of 
$T$, that is, the collection of all ground atoms $p(c_1,\ldots, c_n)$, 
where $p$ is an $n$-ary relation symbol appearing in $T$ and $c_i\in 
\HU(T)$, $1\leq i\leq n$. Following a standard practice, we identify 
Herbrand interpretations of $T$ with subsets of $\HB(T)$.

The restriction to Herbrand interpretations is important. In
particular, it implies that the logic $\plc$ is {\em nonmonotonic}. 
Indeed, if $T_1 \subseteq T_2$ are two $\plc$ theories, it is not 
necessary that every Herbrand model of $T_2$ is a Herbrand model
of $T_1$. For example, let $T_1 = \{\top \rra p(\und),\ \
p(a)\rra\bot\}$, and let $T_2 = T_1 \cup \{\top \rra p(b)\}$. It is 
easy to see that $M=\{p(b)\}$ is a Herbrand model of $T_2$ and that
$T_1$ has no Herbrand models. Indeed, $\HU(T_1)=\{a\}$ and the only
Herbrand model satisfying the first rule is $M'=\{p(a)\}$. This model,
however, does not satisfy the second rule. In contrast, classical
first-order logic is monotone: for every two collections of sentences 
$T_1\subseteq T_2$, if $M$ is a model of $T_2$ then it is a model of 
$T_1$, as well. In the example discussed above, the Herbrand model 
specified by the subset $\{p(b)\}$ of $\HB(T_2)$ is a model of $T_1$ 
but {\em not} a Herbrand model of $T_1$.

The restriction to Herbrand models allows us also to develop algorithms
to compute them. Namely, as in the case of stable logic programming, 
Herbrand models of a $\plc$ theory $T$ can be computed in two steps. 
First, we {\em ground} $T$ to a propositional theory that has the same
Herbrand models as $T$. Next, we compute models of $T$ by computing 
models of the ground theory. This latter task can be accomplished by
off-the-shelf propositional satisfiability solvers. 

The concept of 
grounding is similar to that used in the context of universal theories 
in first-order logic or programs in logic programming. The only 
difference comes from the fact that rules in $\plc$ theories may 
include e-atoms in the consequents. 
We will now discuss the task of grounding in detail.
Let $p(t)$ be an atom that occurs in $T$. If $p(t)$ is not an e-atom, 
we define $p^d(t)=p(t)$. If $p(t)$ is an e-atom (we assume that all 
e-atoms in $T$ are given in the ``underscore'' notation), we define
$p^d(t)$ as the disjunction of all atoms of the form $p(t')$, where 
$t'$ is obtained from $t$ by replacing all occurrences of the 
underscore symbol $\und$ in $t$ with constants from $\HU(T)$ (that 
is, constants that appear in $T$). For example, if $a$ and $b$ are 
the only two constants in $T$ and $p(\und,X,\und,a)$ is an e-atom 
in $T$, we have
\[
p^d(\und,X,\und,a) = p(a,X,a,a) \vee p(a,X,b,a) \vee p(b,X,a,a)
\vee p(b,X,b,a).
\]

Further, for a rule $r\in T$, where 
\[
r = \ \ A_1 \wedge \ldots A_m \rra B_1 \vee\ldots \vee B_n,
\]
we define
\[
r^d = \ \ A_1 \wedge \ldots A_m \rra B^d_1 \vee\ldots \vee B^d_n.
\]
We note that $T^d=\{r^d\colon r\in T\}$ contains no occurrences of 
the underscore symbol. 

Let $\vartheta$ be a {\em ground substitution}, that is, a mapping 
whose domain is a finite subset of the set of variables from the 
language, and which assigns constants from the language to variables 
in its domain. By an {\em expression} we mean a term, a list of terms, 
a formula without any occurrence of the underscore symbol, or a set 
of formulas without any occurrence of the underscore symbol. A ground 
substitution is {\em defined} for an expression $E$ if every variable 
appearing in $E$ belongs to the domain of $\vartheta$. Let $E$ be an 
expression and let $\vartheta$ be a ground substitution that is 
defined for $E$. By $E\vartheta$ we denote the expression obtained 
from $E$ by replacing all variables occurring in $E$ with their images 
under $\vartheta$. We call an expression of the form $E\vartheta$ a 
{\em ground instance} of $E$. For an expression $E$, by $\gr(E)$ we 
denote the set of all ground instances of $E$. Finally, we extend the
notion of grounding to $\plc$ theories which are not, in general,
expressions as they may contain the underscore symbol. Namely, for
$\plc$ theory $T$, we set $\gr(T) = \gr(\{r^d\colon r\in T\})$. 

We will illustrate the concepts we introduced with an example. 
Let $T$ be a  $\plc$ theory that consists of the following two 
clauses:
\begin{quote}
$C_1 =\ \ q(b,c) \rra p(a)$\\
$C_2 =\ \ p(X)\rra q(X,\und)$.
\end{quote}
To compute $\gr(T)$ we need to compute all ground instances of $C_2$
($C_1$ is already in the ground form). First, we compute
the formula $C_2^d$:
\[
C_2^d = \ \ p(X)\rra q(X,a)\vee q(X,b)\vee q(X,c).
\]
To obtain all ground instances of $C_2$ (or $C_2^d$), we replace $X$
in $C_2^d$ with $a$, $b$ and $c$ and obtain the following three clauses:
\begin{quote}
$p(a)\rra q(a,a)\vee q(a,b)\vee q(a,c)$\\
$p(b)\rra q(b,a)\vee q(b,b)\vee q(b,c)$\\
$p(c)\rra q(c,a)\vee q(c,b)\vee q(c,c)$.
\end{quote}
These three clauses together with $C_1$ form $\gr(T)$. 

The following proposition establishes the adequacy of the concept of 
grounding in the study of models of $\plc$ theories. It also demonstrates
that satisfiability provers can be used to compute models of $\plc$ 
theories. The proof of the proposition is simple and reflects closely 
the corresponding argument in the first-order case. Thus, we omit it.

\begin{proposition}
Let $T$ be a $\plc$ theory. A set $M\subseteq \HB(T)$ is a Herbrand
model of $T$ if and only if $M$ is a propositional model of $\gr(T)$. 
\end{proposition}

\section{Equality and arithmetic in the logic $\plc$}

In the following sections, we will often consider a version of the 
logic $\plc$ in which some relation symbols are given a prespecified
interpretation. They are quality, inequality and basic arithmetic relations 
such as $\leq$, $<$, $\geq$, $>$, $+$, $*$, $-$, $/$, etc. Inclusion
of these relation symbols in the language is important as they greatly 
facilitate the task
of programming (modeling knowledge, representing constraints as rules). 
We will use standard symbols to represent them, as well as the standard 
infix notation. In particular, we will write $t_1=t_2$ rather than 
$=(t_1, t_2)$, and $t=t_1+t_k$ (or $t_1+ t_2 =t$) rather than 
$+(t_1,t_2,t)$. We will denote the set of these relation symbols by 
$EA$.

In this section, we will define the semantics for this variant 
of the logic $\plc$. The idea is to interpret all symbols in the set 
$EA$ according to their intended meaning. Specifically, let $C$ be the 
set of constant symbols. We define a theory $=_C$ to 
consist of all clauses of the form
\begin{enumerate}
\item $\top \rra =(t,t)$ (we will write them as $\top \rra (t=t)$), for 
every $t\in C$, and
\item $=(s,t)\rra \bot$ (we will write them as $(s=t) \rra \bot$), 
for all $s,t\in C$ such that $s\not=t$.
\end{enumerate}
Next, we define a theory $+_C$ to consist of all clauses of the form
\begin{enumerate}
\item $\top \rra +(t,u,s)$ (we will write them as $\top \rra (s=t+u)$), 
for every integers $s,t,u \in C$ such that $s=t+u$, and
\item $+(t,u,s) \rra \bot$ (we will write them as $(s=t+u) \rra \bot$), 
for every $s,t,u\in C$ such that at least one of $s,t,u$ is not an 
integer, or $s,t,u$ are integers and $s \not= t+u$.
\end{enumerate}
In the same way we define theories $p_C$ for other relation symbols in 
$EA$ such as $\leq$, $-$, $*$, etc. All these theories provide explicit
intended definitions of the corresponding relation symbols. We will 
often refer to the relation symbols in $EA$ as {\em predefined} since 
their interpretation is fixed.

Let $T$ be a $\plc$ theory in the language containing distinguished
relation symbols from the set $EA$. Let $C$ be the set of constants
appearing in $T$ (that is, $C= \HU(T)$). A set $M$ of ground atoms in 
the language is a {\em model} of $T$ if $M$ is a model of the theory 
$T\cup \bigcup \{p_C \colon p\in EA\}$ as defined above.

It is clear that, with the help of additional variables, we can
express in the logic $\plc$ arbitrary arithmetic expressions. For 
instance, we will write
\[
q((X+Y)*Z,a)\wedge B \rra H
\]
and interpret this expression as
\[
(T_1=X+Y)\wedge (T_2=T_1*Z)\wedge q(T_2,a) \wedge B \rra H
\]
Similarly, we will interpret
\[
B \rra q((X+Y)*Z,a)\vee H
\]
as the clause
\[
B\wedge (T_1=X+Y)\wedge (T_2=T_1*Z) \rra q(T_2,a) \vee H.
\]
In each case, variables $T_1$ and $T_2$ are different from all other 
variables appearing in the rules. In order to obtain uniqueness of 
the interpretation, when decomposing arithmetic expressions, we 
follow the standard order in which they are evaluated. Let us emphasize
that arithmetic expressions are simply notational shortcuts and not elements of the language.

In the remainder of the paper, we will always assume that the language
contains predefined relation symbols. Since their extensions are 
already fully specified, we will omit the corresponding ground atoms
when describing models of theories.

To illustrate these concepts, let us consider the following $\plc$ 
theory $T$:
\[
T=\{\top \rra p(1),\ \ \top \rra p(2),\ \ p(X) \rra q(X,X+1)\}.
\]
This theory represents the following $\plc$ theory $T'$:
\[
T'=\{\top \rra p(1),\ \ \top \rra p(2),\ \ p(X)\wedge (Y=X+1) \rra
q(X,Y)\}.
\]
The theory $\gr(T')$ consists of the first two rules (they are already
ground) and the following four instantiations of the third rule:

\begin{quote}
$p(1)\wedge (1=1+1) \rra q(1,1)$\\
$p(1)\wedge (2=1+1) \rra q(1,2)$\\
$p(2)\wedge (1=2+1) \rra q(2,1)$\\
$p(2)\wedge (2=2+2) \rra q(2,2)$.
\end{quote}

Models of this theory are (by the definition) models of the theory $T$.
For instance, $\{p(1), p(2), q(1,2)\}$ is a model of $T$. We point out
that, according to our convention, we omitted from the model description
the atom $2=1+1$ (or, more formally, the atom $=(1,1,2)$).

\section{Programming with logic $\plc$}

The logic $\plc$, described in the previous section, can be used as a 
basis for a declarative programming formalism based on the paradigm 
of answer-set programming \cite{mt99}. To this end, we need to
introduce the concepts of input data and a program. We follow the 
approach proposed and studied in the area of relational databases 
\cite{ul88}. A relational database can be viewed as a collection of 
ground atoms of some logic language. We often use (for instance, in the
context of DATALOG and its variants) the term {\em extensional} 
database to refer to a collection of ground atoms specifying a
relational database. Queries are finite theories, often of special 
form, in this logic language (for instance, definite Horn theories 
without function symbols serve as queries in the case of DATALOG).
Queries define new properties (relations) in terms of those relations 
that are explicitly specified by the underlying (extensional) database. 

Guided by these intuitions, we define a {\em data-program pair} to be a
pair $(D,P)$, where $D$ is a finite set of ground atoms in a language
of the logic $\plc$ and $P$ is a finite collection of $\plc$ rules. We 
use data-program pairs to represent specific computational 
problem instances. We view $D$ as an encoding of relevant input data 
and $P$ as a declarative specification of the computational task in 
question. Accordingly, given a data-program pair $(D,P)$, we refer to 
$D$ as a {\em data set} and to $P$ as a {\em program}. We use the 
term {\em data predicate} for all relation symbols appearing in atoms 
in $D$. We use the term {\em program predicate} to refer to all 
relation symbols that appear in $P$ and are neither data predicates 
nor predefined predicates from $EA$. Intuitively, $D$ is a counterpart 
of an extensional database and $P$ is a counterpart of a database query. 
 
We will now introduce a semantics for data-program pairs. To this 
end, we will encode a data-program pair $(D,P)$ as a theory in the 
logic $\plc$. Since $D$ is a set of ground atoms representing the 
problem instance (input data), we assume that $D$ provides a {\em 
complete} specification of the input. That is, we assume that no other 
ground atoms built of predicates appearing in $D$ are true. Since the 
only formulas in the logic $\plc$ are rules, we encode the information
specified by $D$ as a set of rules $\cl(D)$ defined as follows. For 
every relation symbol $p$ appearing in $D$ and for every ground tuple 
$t$ (with constants from the Herbrand universe of $D\cup P$) of 
appropriate arity, if $p(t)\in D$, we include in $\cl(D)$ the clause
\[
\top \rra p(t).
\]
Otherwise, if $p(t)\notin D$, we include in $\cl(D)$ the clause
\[
p(t)\rra \bot.
\]
It is clear that the set $\cl(D)$ can be regarded as the result of 
applying Reiter's Closed-World Assumption to $D$.

We represent a data-program pair $(D,P)$ by a $\plc$ theory $\cl(D)
\cup P$. We say that a set $M$ of ground atoms, $M\subseteq \HB(\cl(D)
\cup P)$, is a {\em model} of a data-program pair $(D,P)$ if it is a 
model of $\cl(D)\cup P$. We denote the set of all models of a
data-program pair $(D,P)$ by $\mod(D,P)$.

In separating data and program predicates and in adopting the 
closed-world assumption for the treatment of data atoms we are 
guided by the intuition that data predicates are intended to 
represent input data. Their extensions should not be affected by the 
computation. The effects of the computation should be reflected in 
the extensions of program predicates only. 

We designed the logic $\plc$ and introduced the concept of a
data-program pair to model computational problems. 
To illustrate this use of our formalism, we show how to encode several 
well-known search problems by means of data-program pairs. We assume 
that the language contains predefined relation symbols to represent 
equality and arithmetic relations. 

We start with the graph $k$-colorability problem: given an 
undirected graph and a set of $k$ colors, the objective is to find an 
assignment of colors to vertices so that no two identically colored 
vertices are joined with an edge (or to determine that no such coloring 
exists). 

We set
\[
D_\gcl(G,k) = \{\vtx(v)\colon v\in V\} \cup \{\edge(v,w)\colon \{v,w\}\in
E\} \cup \{\col(i)\colon 1\leq i\leq k\}.
\]
The set of atoms $D_\gcl$ represents an instance of the coloring problem.
The predicates $\vtx$, $\edge$ and $\col$ are data predicates. Their
extensions define vertices and edges of an input graph, and the set of 
available colors. 

Next, we construct a program, $P_\gcl$, encoding the constraints of 
the problem. It involves predicates $\vtx$, $\edge$ and $\col$, 
specified in the data part, and defines a new relation $\clr$ that 
models assignments of colors to vertices.

\newcounter{ct2}
\begin{list}{C\arabic{ct2}:\ }{\usecounter{ct2}\topsep 0.03in
\parsep 0in\itemsep 0in}
\item $\clr(X,C)\rra \vtx(X)$
\item $\clr(X,C)\rra \col(C)$
\item $\vtx(X)\rra \clr(X,\und)$
\item $\clr(X,C)\wedge\clr(X,D) \rra (C=D)$
\item $\edge(X,Y)\wedge \clr(X,C) \wedge \clr(Y,C) \rra \bot$.
\end{list}

The condition (C1) states that the only objects that get colored are 
vertices. Indeed, by the definition, a model of the theory 
$(D_\gcl(G,k), P_\gcl)$ contains an atom $\vtx(x)$ if and
only if $x$ is a vertex of an input graph. Thus, if $\clr(v,c)$ belongs
to a model of $(D_\gcl(G,k), P_\gcl)$, then $\vtx(v)$ belongs to the
model and, so, $v$ is a vertex. Similarly, (C2) states that 
the only objects assigned by the predicate $\clr$ to a vertex are colors. 
(C3) states that each vertex $X$ gets assigned at least one color. (C4) 
enforces that each vertex is assigned at most one color. (C5) ensures 
that two vertices connected by an edge are assigned different colors. 
These clauses correctly capture the constraints of the coloring 
problem. 

\begin{proposition}
\label{p-col}
Let $G=(V,E)$ be an undirected graph and let $k$ be a positive integer.
An assignment $f\colon V \rightarrow \{1,\ldots,k\}$ is a $k$-coloring 
of $G$ if and only if $M = D_\gcl(G,k) \cup \{\clr(v,f(v))\colon v\in V\}$ 
is a model of the data-program pair $(D_\gcl(G,k),P_\gcl)$. 
\end{proposition}
Proof: $(\Ra)$ Let us assume that $f\colon V \rightarrow
\{1,\ldots,k\}$ is a $k$-coloring of $G$. We will show that $M =
D_\gcl(G,k) \cup \{\clr(v,f(v))\colon v\in V\}$ is a model of 
$(D_\gcl,P_\gcl)$, that is, it is a model of $\gr(\cl(D_\gcl(G,k))\cup
P_\gcl)$.  From the definition of $M$ it follows that M satisfies all 
rules in $\cl(D_\gcl(G,k))$. We will now show that $M$ satisfies all 
rules in $\gr(\cl(D_\gcl(G,k))\cup P_\gcl)$ that are obtained by 
grounding rules in $P_\gcl$. 

First, we consider an arbitrary ground instance of rule (C1),
say, $\clr(x,c)\rra \vtx(x)$, where $x$ and $c$ are two constants 
of the language. It is clear from the definition of $M$ that if 
$\clr(x,c) \in M$, then $x\in V$ and, consequently, $\vtx(x)\in M$. 
Thus all ground instances of (C1) are satisfied by $M$.

Next, we consider a ground instance $r$ of rule (C3), say,
\[
r =\ \ \vtx(x)\Rightarrow \bigvee\{\clr(x,c)\colon c\in V\cup
\{1,\ldots,k\}\},
\]
where $x\in V\cup \{1,\ldots,k\}$. If $\vtx(x)\in
M$, then $x\in V$. Since $f(x)\in \{1,\ldots,k\}$, and $\clr(x,f(x))
\in M$, it follows that $r$ is satisfied by $M$. All other rules can 
be dealt with in a similar way.
\ \\
$(\La)$ We will now assume that $M$ is a model of $(D_\gcl,P_\gcl)$.
By the definition of a model, we have (1) $\vtx(x)\in M$ if and only 
if $x\in V$, (2) $\edge(x,y)\in M$ if and only if $\{x,y\}\in E$, and 
(3) $\col(i)\in M$ if and only if $i\in \{1,\ldots,k\}$.

Now, we observe that since $M$ satisfies all ground instances of (C1),
if $\clr(x,c)\in M$, then $x\in V$. Similarly, since $M$ satisfies all 
ground instances of (C3), for every $x\in V$ there is at least
one constant $c$ such that $\clr(x,c)\in M$. On the other hand, since
$M$ satisfies all ground instances of (C2), for each such constant $c$,
$c\in \{1,\ldots,k\}$. Next, we have that $M$ satisfies all ground
instances of (C4). Consequently, for every $x\in V$ there is exactly
one $c\in \{1,\ldots,k\}$ such that $\clr(x,c)\in M$. Let us denote by
$f$ the function that assigns to each $x\in V$ the unique $c\in\{1,
\ldots,k\}$ such that $\clr(x,c)\in M$. It follows that $M = D \cup
\{clr(v,f(v))\colon v\in V\}$. Moreover, as $M$ satisfies all ground
instances of (C5), $f$ is a $k$-coloring of $G$. \hfill $\Box$

Let us note that the proof of Proposition \ref{p-col} implies, in fact, that 
the correspondence between models and colorings is a bijection.

Next, we will describe a data-program pair encoding an instance of the 
vertex-cover problem for graphs. Let $G=(V,E)$ be a graph. A set 
$W\subseteq V$ is a {\em vertex cover} of $G$ if for every edge 
$\{x,y\}\in E$, $x$ or $y$ (or both) are in $W$. The vertex-cover 
problem is defined as follows: given a graph $G=(V,E)$ and an integer 
$k$, $k\leq |V|$, decide whether $G$ has a vertex cover with no more 
than $k$ vertices. 

For the vertex-cover problem the input data is described by the
following set of ground atoms: 
\[
D_\vc(G,k) = \{\vtx(v)\colon v\in V\} \cup \{\edge(v,w)\colon \{v,w\}\in
E\} \cup \{\indx(i)\colon i=1,\ldots, k\}.
\]
This set of atoms specifies the set of vertices and the set of edges 
of an input graph. It also provides a set of $k$ {\em indices} which 
we will use to select a subset of no more than $k$ vertices in the graph,
a candidate for a vertex cover of cardinality at most $k$. 

The vertex cover problem itself is described by the program $P_\vc$. It 
introduces a new relation symbol $\vc$. Intuitively, we use $\vc$ to
represent the fact that a vertex has been selected to a candidate set. 

\newcounter{ct22}
\begin{list}{VC\arabic{ct22}:\ }{\usecounter{ct22}\topsep 0.03in
\parsep 0in\itemsep 0in}
\item $\vc(I,X) \rra \vtx(X)$
\item $\vc(I,X)\rra \indx(I)$
\item $\indx(I)\rra \vc(I,\und)$
\item $\vc(I,X) \wedge \vc(I,Y) \rra X=Y$ 
\item $\edge(X,Y) \rra \vc(\und,X) \vee \vc(\und,Y)$.
\end{list}

(VC1) and (VC2) ensure that $\vc(i,x)$ is false if $i$ is not an 
integer from the set $\{1,\ldots, k\}$ or if $x$ is not a vertex (that
is, if $\vc(i,x)$ is true, $i\in \{1,\ldots,k\}$ and $x\in V$). The 
rules (VC3) and (VC4) together impose the requirement that every index 
$i$ has exactly one vertex assigned to it. It follows that the set of 
ground atoms $\vc(i,x)$ that are true in a model of the data-program 
pair $(D_\vc(G,k), P_\vc)$ defines a subset of $V$ with cardinality at 
most $k$. Finally, (VC5) ensures that each edge has at least one end
vertex assigned by $\vc$ to an index from $\{1,\ldots,k\}$ (in other 
words, that vertices assigned to indices $1,\ldots, k$ form a vertex 
cover). The correctness of this encoding is formally established in the 
following result. Its proof is similar to that of Proposition \ref{p-col} 
and we omit it.

\begin{proposition}
\label{p-vtx}
Let $G=(V,E)$ be an undirected graph and let $k$ be a positive integer.
If $W\subseteq V$ is a vertex cover of $G$ and $|W|\leq k$, then for 
every sequence $w_1,\ldots,w_k$ that enumerates all elements in $W$ 
(possibly with repetitions), $M = D_\vc(G,k) \cup \{\vc(i,w_i)\colon 
i=1,\ldots, k\}$ is a model of the data-program pair $(D_\vc(G,k),
P_\vc)$. Conversely, if $M$ is a model of $(D_\vc(G,k),P_\vc)$ then
the set $W=\{w\in V\colon \vc(i,w)\in M,\ \mbox{for some $i=1,\ldots,
k$}\}$ is a vertex cover of $G$ with $|W|\leq k$.
\end{proposition}

In this case, we do not have a one-to-one correspondence between models 
and vertex covers of cardinality at most $k$. It is so because we 
represent sets by means of sequences.

Next, we will consider the Hamiltonian-cycle problem in a {\em directed}
graph. To represent an input graph $G=(V,E)$ we use the following set of 
ground atoms:
\[
D_\hc(G) = \{\vtx(x)\colon x\in V\} \cup \{\edge(x,y)\colon (x,y)\in E\}
\cup \{\indx(i)\colon i=1,\ldots,|V|\}.
\]
The set of indices is introduced as part of input because we will 
represent a Hamiltonian cycle by a bijective sequence of vertices such 
that every two consecutive vertices in the sequence. as well as the last
and the first, are connected with an edge. To represent such sequences
we use a relations symbol $\hcp$. The program, $P_\hc$, defining 
``Hamiltonian'' sequences $\hcp(i,x)$ looks as follows. In this example
we assume that $\oplus$ denotes a predefined relation of addition modulo 
$n$ defined on the set of integers $\{1,\ldots, n\}$ (thus, in particular, 
$n\oplus 1=1$).

\newcounter{ct23}
\begin{list}{HC\arabic{ct23}:\ }{\usecounter{ct23}\topsep 0.03in
\parsep 0in\itemsep 0in}
\item $\hcp(I,X) \rra \indx(I)$
\item $\hcp(I,X) \rra \vtx(X)$
\item $\indx(I) \rra \hcp(I,\und)$
\item $\hcp(I,X) \wedge \hcp(I,Y) \rra X=Y$
\item $\hcp(I,X) \wedge \hcp(J,X) \rra I=J$
\item $\hcp(I,X)\wedge \hcp(I \oplus 1,Y) \rra edge(X,Y)$.
\end{list}
The first two rules ensure that if $\hcp(i,x)$ is true in a model of
$(D_\hc,P_\hc)$ then $i$ is an integer from the set $\{1,\ldots,|V|\}$
and $x\in V$. The rules (HC3) - (HC5) together enforce the constraint 
that $\hcp$ defines a {\em permutation} of vertices. Finally, the last 
rule imposes the Hamiltonicity constraint that from every vertex in the
sequence to the next one (and from the last one to the first one, too)
there is an edge in the graph. Formally, we have the following result
(the correspondence it establishes is not one to one as a Hamiltonian 
cycle can be represented by $|V|$ different permutations, each being a 
cyclic shift of another).

\begin{proposition}
\label{h-vtx}
Let $G=(V,E)$ be a directed graph with $n$ vertices. A permutation $v_1,
\ldots, v_n$ of all vertices of $V$ is a Hamiltonian cycle if and only
if $M = D_\hc(G) \cup \{\hcp(i,v_i)\colon i=1,\ldots, n\}$ is a model 
of the data-program pair $(D_\hc(G), P_\hc)$. 
\end{proposition}

We will next consider the $n$-queens problem, that is,
the problem of placing $n$ queens on a $n\times n$ chess board so that
no queen attacks another. The representation of input data specifies the 
set of row and column indices:
\[
D_\nq(n) = \{\indx(i)\colon i=1,\ldots, n\}.
\]
The problem itself is described by the program $P_\nq$. The predicate
$q$ describes a distribution of queens on the board: $q(x,y)$ is true
precisely when there is a queen in the position $(x,y)$.

\newcounter{ct17}
\begin{list}{nQ\arabic{ct17}:\ }{\usecounter{ct17}\topsep 0.03in
\parsep 0in\itemsep 0in}
\item $q(R,C) \rra \indx(R)$
\item $q(R,C) \rra \indx(C)$
\item $\indx(R) \rra q(R,\und)$
\item $q(R,C1)\wedge q(R,C2) \rra (C1=C2)$
\item $q(R1,C)\wedge q(R2,C) \rra (R1=R2)$
\item $q(R,C) \wedge q(R+I,C+I) \rra \bot$
\item $q(R,C) \wedge q(R+I,C-I) \rra \bot$
\end{list}

The first two rules ensure that if $q(r,c)$ is true in a model of 
$(D_\nq,P_\nq)$ then $r$ and $c$ are integers from the set $\{1,\ldots,
n\}$. The rules (nQ3) - (nQ5) together enforce the constraint that each 
row and each column contains exactly one queen. Finally, the last two 
rules guarantee that no two queens are placed on the same diagonal. As 
in the other cases, we can formally state and prove the correctness of 
this encoding. The proof is again quite similar to that of Proposition 
\ref{p-col} and so we we omit it.

\begin{proposition}
\label{nqueens}
Let $n$ be a positive integer. A set of points on an $n\times n$ board,
$\{(r_i,c_i)\colon i=1,2,\ldots,n\}$, is a solution to the $n$-queens 
problem if and only if the set $M= D_\nq(n)\cup \{q(r_i,c_i)\colon i=1,2,
\ldots,n\}$ is a model of the data-program pair $(D_\nq(n),P_\nq)$.
\end{proposition}

As in the case of the graph-coloring problem, the correspondence 
between models and valid arrangements of queens on the board is 
a bijection.

For the last example in this section, we consider computing the 
transitive closure of a finite directed graph $G =(V,E)$, where $V$ 
is a set of vertices and $E$ is a set of directed edges (we will
assume that $G$ has no loops). We recall 
that the transitive closure of the graph $G=(V,E)$ is the directed 
graph $(V,E')$ such that an edge $(x,y)$ belongs to $E'$ if and only 
if there is in $G$ a directed path from $x$ to $y$ of length at least
1. 

We will now describe the representation of data instances and give
a $\plc$ program solving the transitive closure problem. The data
instance consists of a specification of an input graph $(V,E)$ and
of a collection of integers, $\{1, 2,\ldots, |V|\}$ that will allow us
to count edges in the paths. Thus, we set

\[
D_\tc(G) = \{\vtx(v)\colon v\in V\} \cup \{\edge(v,w)\colon
\{v,w\}\in E\} \cup \{\indx(i)\colon 1\leq i\leq k\}.
\]

Next, we construct a program, $P_\tc$, encoding the constraints 
of the problem. Our encoding uses an auxiliary 4-ary relation 
symbol $\pth$. The intended meaning of $\pth(X,Y,Z,I)$ is that 
it is true precisely when there is a directed path from $X$ to $Y$ 
such that $Z$ is the immediate predecessor of $Y$ on the path and 
the path length is at most $I$. In $P_\tc$ we define the relation 
$\pth$ and use it to specify the relation $\tc$ that represents 
the transitive closure of the input graph.

\setcounter{ct2}{0}
\begin{list}{TC\arabic{ct2}:\ }{\usecounter{ct2}\topsep 0.03in
\parsep 0in\itemsep 0in}
\item $\pth(X,Y,Z,I) \rra \vtx(X)$
\item $\pth(X,Y,Z,I) \rra \vtx(Y)$
\item $\pth(X,Y,Z,I) \rra \vtx(Z)$
\item $\pth(X,Y,Z,I) \rra \indx(I)$
\item $\tc(X,Y) \rra \vtx(X)$
\item $\tc(X,Y) \rra \vtx(Y)$
\item $\pth(X,Y,X,1) \rra \edge(X,Y)$
\item $\edge(X,Y) \rra \pth(X,Y,X,1)$
\item $\pth(X,Y,Z,1) \rra X=Z$
\item $\pth(X,Y,Z,I+1) \rra \pth(X,Z,\und,I)$
\item $\pth(X,Y,Z,I+1) \rra \edge(Z,Y)$
\item $\pth(X,Z,W,I) \wedge e(Z,Y) \rra \pth(X,Y,Z,I+1)$
\item $\tc(X,Y) \rra \pth(X,Y,\und,\und)$
\item $\pth(X,Y,Z,I) \rra \tc(X,Y)$.
\end{list}

The first four rules enforce that if an atom $\pth(x,y,z,i)$ is in 
a model of the data-program pair $(D_\tc(G),P_\tc)$ then $x$, $y$ and 
$z$ are vertices ($\vtx(x)$, $\vtx(y)$ and $\vtx(z)$ hold) and $i$ is 
an index ($\indx(i)$ holds). The effect of the next two rules is 
similar but they concern the relation symbol $\tc$. The rules (TC7) 
- (TC9) enforce conditions that atoms $\pth(x,y,z,1)$ must satisfy 
to be in a model. The rules (TC10) - (TC12) enforce recursive 
conditions that atoms $\pth(x,y,z,i)$, $i\geq 2$, must satisfy in 
order to be in the model. Finally, the rules (TC13) - (TC14) define 
the relation symbol $\tc$ in terms of the relation $\pth$.

The following result can now be proved by an easy induction.

\begin{proposition}
\label{tc}
Let $G$ be a directed graph. The data-program pair $(D_\tc(G),P_\tc)$
has a unique model that consists of (1) all atoms in $D_\tc(G)$, (2)
all atoms $\pth(x,y,z,i)$ such that there is a path in $G$ from $x$ 
to $y$ of length $i$ and with $z$ being the last but one vertex on 
this path, and (3) of all atoms $\tc(x,y)$ such that there is a 
directed path of positive length from $x$ to $y$ in $G$.
\end{proposition}

We have chosen to discuss in detail the question of the transitive
closure since it is well known that this property is not definable in 
first-order logic \cite{av91}. We can define it in our logic $\plc$ 
because our notion of definability is different: data-program pairs
define concepts as special Herbrand models. A more detailed discussion
of these issues follows in the next section. 

\section{Expressive power of the logic $\plc$}
\label{eps}

Our discussion in the previous sections demonstrated the use of the 
logic $\plc$ as a tool to represent computational problems.
In this 
section, we will study the expressive power of the logic $\plc$, that
is, we will identify a class of computational problems that can be 
represented by means of finite $\plc$ programs. 

We first recall some database terminology \cite{ul88}. Let $\dom$ 
be a fixed infinite set (for instance, the set of all natural numbers). 
A {\em relational schema} over a domain $\dom$ is a nonempty sequence 
$R=(r_1, \ldots, r_k)$ of relation symbols. Each relation symbol $r_i$ 
comes with integer arity $a_i>0$. An {\em instance} of a 
relation schema $R$ is a {\em nonempty} and {\em finite} set of ground 
atoms, each of the form $r_i(u_1,\ldots, u_{a_i})$, where $1 \leq 
i\leq k$ and $u_1,\ldots, u_{a_i}\in \dom$. By
${\cal I}(R)$ we denote the set of all instances of a relational schema 
$R$. Since $\dom$ is fixed, form now on we will not explicitly mention 
it. We also emphasize that, unlike in standard presentations, we
require that instances of a relation schema be nonempty.

Relational schemas provide a framework for a precise definition
to a class of computational problems known as {\em search problems}.
Let $R$ and $S$ be two disjoint relational schemas. A {\em search
problem} (over relational schemas $R$ and $S$) is a recursive relation 
$\Pi\subseteq {\cal I}(R)\times {\cal I}(S)$. The set ${\cal I}(R)$ is 
the set of {\em instances} of $\Pi$. Given an instance $I\in {\cal 
I}(R)$, the set $\{J\in {\cal I}(S): (I,J)\in \Pi\}$ is 
the set of {\em solutions} to $\Pi$ for the instance $I$. 

Search problems abound. It is clear that the graph problems and the 
$n$-queens problem considered earlier in the paper are examples of 
search problems. More generally, all constraint satisfaction problems 
over discrete domains, including such basic AI problems as planning, 
scheduling and product configuration, can be cast as search problems.

A {\em search language} or ({\em language} for short) is a set $L$ of 
expressions and a function $\mu$ such that for every expression $e\in L$,
$\mu(e)$ is a search problem. We call $\mu$ the {\em interpretation
function} for $L$. By the {\em expressive power} of a language $L$
we mean the class of search problems defined by expressions from $L$: 
$\{\mu(e) \colon e\in L\}$. 

We note that the concept of a search problem extends that of a database 
{\em query} \cite{var82}, which is defined as a partial recursive 
{\em function} from ${\cal I}(R)$ to ${\cal I}(S)$. Consequently,
fragments of search languages consisting of those expressions that 
define partial functions are, in particular, database query languages. 
In fact, one can regard a search problem as a {\em second-order query} 
--- a mapping from the set of instances of some relational schema $R$ 
into the {\em power set} of the set of instances of another (disjoint) 
relational schema $S$. Pushing the analogy further, a search language 
can be viewed as a second-order database query language --- an 
expression in such a language defines, given an instance of a
relational schema $R$, a collection of instances of a relation schema
$S$ rather than a single instance.

We will show that the logic $\plc$ gives rise to a search language 
and establish its expressive power. An {\em expression} is a pair
$(P,R,S)$, where $P$ is a $\plc$ program, and $R$ and $S$ are disjoint
nonempty sets of relation symbols in $P$. 
We will show that $(P,R,S)$ can be viewed as a specification of 
a search problem over relational schemas $R$ and $S$. Namely, 
let $D\in {\cal I}(R)$. For every set $M\subseteq 
\HB(D\cup P)$, by $M[S]$ we denote the set of all those atoms in $M$ 
that are built by means of relation symbols from $S$. We define the 
interpretation function $\mu$ as follows:
\[
\mu(P,R,S) = \{(D,F)\colon D\in {\cal I}(R),\ \mbox{and}\ F = M[S],\ \
\mbox{where $M\in \mod(D,P)$}\}.
\]
It is clear that $\mu(P,R,S) \subseteq {\cal I}(R)\times{\cal I}(S)$.
Consequently, the set of $\plc$ expressions together with the function
$\mu$ is a search language.

In a similar way we can view as a search language the language of 
$\datan$ (logic programming without function symbols) with the 
semantics of Herbrand models, supported models \cite{cl78,ap90} or 
stable models \cite{gl88}. Since the expressive power of $\datan$ with 
the supported-model semantics will play a role in our considerations, we 
will recall relevant notions and results\footnote{All concepts related 
to $\datan$ that we mention here can be defined in a more general 
setting of logic programming languages that include function symbols. 
For an in-depth discussion of logic programming, we refer the reader 
to \cite{ap90}.}. 

Let $\cal L$ be a language of predicate logic. A $\datan$ {\em clause} 
is an expression $r$ of the form
\[
r =\ \ \ p(X) \lla q_1(X_1),\ldots, q_m(X_m),
           \neg q_{m+1}(X_{m+1}),\ldots, \neg q_{m+n}(X_{m+n}),
\]
where $p, q_1, \ldots, q_{m+n}$ are relation symbols and $X,X_1,\ldots, 
X_{m+n}$ are tuples of constant and variable symbols with arities 
matching the arities of the corresponding relation symbols. We call the 
atom $p(X)$ the {\em head} of the clause $r$ and denote it by $h(r)$. If 
a clause has empty body, we represent it by its head (thus, atoms can be 
regarded as clauses). For a clause $r$ we also set
\[
B(r) = q_1(X_1) \wedge \ldots \wedge q_m(X_m) \wedge \neg 
q_{m+1}(X_{m+1}) \wedge \ldots\wedge \neg q_{n}(X_{n}).
\]

A $\datan$ {\em program} is a collection of $\datan$ clauses. Let $P$ 
be a $\datan$ program. As usual, we call relation symbols that appear 
in the heads of clauses in $P$ {\em intentional}. We refer to all other 
relation symbols in $P$ as {\em extensional}. We denote the sets of 
intentional and extensional relation symbols of a $\datan$ program 
$P$ by $I(P)$ and $E(P)$, respectively. Next, for a relation symbol $p$ 
that appears in $P$, we denote by $\dff(p)$ the set of all clauses in 
$P$ whose head is of the form $p(t)$, for some tuple $t$ of constant
and variable symbols. In other words, $\dff(p)$ consists of all clauses
that {\em define} $p$. 

In the paper we restrict our attention to $\datan$ programs of special
form, called {\em I/O} programs, providing a clear separation of data 
facts (ground atom representing data) from clauses (definitions of 
intentional relation symbols). To this end, we define first a class of 
{\em pure} programs. We say that a $\datan$ program $P$ is {\em pure} 
if 
\begin{enumerate}
\item for every relation symbol $p\in I(P)$, all clauses in $\dff(p)$ 
have the same head of the form $p(X)$, where $X$ is a tuple of distinct 
variables
\item $P$ contains no occurrences of constant symbols
\item $E(P)\not=\emptyset$.
\end{enumerate}
Pure programs are, in particular, in the so-called normal form as they
satisfy condition (1) \cite{ap90}. An {\em I/O program} is a $\datan$ 
program of the form $D\cup P$, where $P$ is a pure program and $D\in
{\cal I}(E(P))$ (that is, $D$ is a nonempty and finite set of ground 
atoms built of relation symbols in $E(P)$). To simplify the discussion, 
we define supported models for I/O programs only. It does not cause any 
loss of generality. Indeed, one can show that for every logic program 
containing at least one constant symbol there is an I/O program with 
the same intentional relation symbols and such that supported models of 
both programs, when restricted to intentional ground atoms, coincide 
(that is, under the semantics of supported models, both programs define 
the same relations).

Let $P$ be a pure program and let $D\in {\cal I}(E(P))$. For a predicate 
$p$ from $I(P)$, we define its {\em (Clark's) completion} $\cc(p)$ as
\[
\cc(p) =\ \ \ p(X)\Leftrightarrow \bigvee \{\exists Y_r\hspace{2mm}
B(r) \colon r\in \dff(p)\},
\]
where $X$ is a tuple of variables and $Y_r$ is the tuple of distinct
variables occurring in the body of $r$ but not in the head of $r$ (we
exploit the normal form of $P$ here) \cite{cl78}. 
We define the (Clark's) completion of $P$, $\CC(P)$, by setting 
\[
\CC(P) = \{\cc(p)\colon p\in \pr\}.
\]
Finally, we define a set of ground atoms $M\subseteq \HB(D\cup P)$ to
be a {\em supported model} of an I/O program $D\cup P$ if it is 
a Herbrand model 
of $\cl(D) \cup \CC(P)$, where $\cl(D)$ is defined as in Section
\ref{plc}. We denote by $\Sup(D\cup P)$ the collection of all supported 
models of $D\cup P$. 

Let $P$ be a pure program and let $S\subseteq I(P)$. We define
\[
\nu(P,S) = \{(D,F)\colon D\in {\cal I}(E(P)),\ \mbox{and}\ F = M[S],
\ \ \mbox{where $D\not=\emptyset$ and $M\in \Sup(D\cup P)$}\}.
\]
Since $E(P)$, $I(P)$ and $S$ can be regarded as relational schemas, 
$\nu(P,S)$ is a search problem. Thus, the set of expressions $(P,S)$, 
where $P$ is a pure program and $S$ is a subset of $I(P)$, together with 
the function $\nu$ form a search language.
 
The expressive power of this language is known. A search problem $\Pi$ 
over relational schemas $R$ and $S$ is in the class {\em NP-search} if 
there is a nondeterministic Turing Machine $\mathit{TM}$ such that 
\begin{enumerate}
\item $\mathit{TM}$ runs in polynomial time
\item for every instance $I$ of the schema $R$ (input instance of $\Pi$),
the set of strings left on the tape when accepting computations for $I$ 
terminate is precisely the set $\{J\in {\cal I}(S)\colon (I,J)\in
\Pi\}$, that is, the set of solutions to $\Pi$ for the input $I$.
\end{enumerate}
The class NP-search is precisely the class of search problems captured 
by finite $\datan$ programs with the supported-model semantics.
\begin{theorem}[\cite{mr00}]
\label{datan}
For every finite pure program $P$ and every $S\subseteq I(P)$, $\nu(P,S)$ 
is a search problem in the class NP-search. Conversely, for every 
problem $\Pi$ in the class NP-search there is a pure program $P$ and
a set $S\subseteq I(P)$ such that $\nu(P,S)=\Pi$.
\end{theorem}

We will now show that the expressive powers of $\plc$ and of $\datan$ 
with supported model semantics are the same. Namely, we will prove the 
following result.

\begin{theorem}
\label{ep-key}
For every finite pure program $P$ and every set $S\subseteq I(P)$, 
there is a finite $\plc$ program $P'$ such that $E(P)\cup I(P)$ are
among the relation symbols appearing in $P'$ and $\nu(P,S)=
\mu(P',E(P),S)$. Conversely, for every finite $\plc$ program $P'$ and 
every nonempty and disjoint sets $R$ and $S$ of relation symbols 
appearing in $P'$, there is a finite pure program $P$ such that
$R=E(P)$, $S\subseteq I(P)$ and $\mu(P',R,S) =\nu(P,S)$.
\end{theorem} 
Proof: Let $P$ be a pure program. We will consider 
the completion $\CC(P)$ of $P$ and construct its equivalent 
representation in terms of $\plc$ rules (we recall that $\plc$ rules 
are just special formulas from the language of predicate logic).

We build this representation of $\CC(P)$ as follows. Let $p$ be a 
predicate symbol in $I(P)$. Let us assume that $p(X)$, where $X$ 
is a tuple of distinct variables, is the common head of all clauses 
in $\dff(p)$. Let us consider a clause $r\in \dff(p)$, say
\[
r = \ \ p(X)\lla q_1(X_1), \ldots, q_m(X_m), \neg 
q_{m+1}(X_{m+1}), \ldots, \neg q_{n}(X_{n}), 
\]
and let $Y_r$ be a tuple of distinct variables that appear in the body 
of $r$ but not in its head. We introduce a new predicate symbol $d_{r}$, 
of the arity $|X|+|Y_r|$ and define the following $\plc$ rules
\begin{quote}
$\psi_i(r) = \ \ d_{r}(X,Y_r) \rra q_i(X_i)$,\ \ $i=1,\ldots,m$\\
$\psi_i(r) = \ \ d_{r}(X,Y_r)\wedge q_i(X_i) \rra \bot$,\ \ $i=m+1,
\ldots,n$\\
$\psi_0(r) = \ \ q_1(X_1) \wedge \ldots \wedge q_m(X_m)
\rra d_{r}(X,Y_r) \vee q_{m+1}(X_{m+1}) \vee \ldots\vee q_{n}(X_{n})$.
\end{quote}
We define $\Psi(r) =\{\psi_0(r),\psi_1(r),\ldots,\psi_n(r)\}$. It is 
clear that $\Psi(r)$ entails (in the first-order logic) the universal 
sentence $d_r(X,Y_r) \leftrightarrow B(r)$ (intuitively, $\Psi(r)$ 
specifies $d_r(X,Y_r)$ so that it can be regarded as an abbreviation 
for $B(r)$).

We will now use atoms $d_r(X,Y_r)$ to define $\plc$ rules that form
an equivalent representation to the formula $\cc(p)$. Let us recall 
that
\[
\cc(p) =\ \ \ p(X)\Leftrightarrow \bigvee \{\exists Y_r\hspace{2mm}
B(r) \colon r\in \dff(p)\}.
\]
Thus, we define the following $\plc$ rules:
\begin{quote}
$\cc'_r(p) = \ \ d_r(X,Y_r) \rra p(X)$,\ \ $r\in \dff(p)$\\
$\cc'(p) = \ \ p(X)\rra \bigvee\{\exists Y_r d_{r}(X,Y_r) \colon r\in 
\dff(p)\}$.
\end{quote}
It is clear (by first-order logic tautologies) that
\[
\Phi(p) = \{\Psi(r)\colon r\in \dff(p)\}\cup \{\cc'_r(p)\colon r\in 
\dff(p)\} \cup \{\cc'(p)\}
\]
and $\cc(p)$ have the same first-order models (modulo new relation 
symbols $d_r$). 

Let us define $P'= \bigcup\{\Phi(p)\colon p\in I(P)\}$. Clearly, $P'$ 
is a $\plc$ program and every relation symbol in $E(P)\cup I(P)$ in 
$P'$. Moreover, by the comment made above, for every instance $D$ 
of the schema $E(P)$, $\cl(D)\cup P$ and $\cl(D) \cup P'$ have the 
same models and, in particular, the same Herbrand models (again modulo 
new relation symbols). Thus, $\nu(P,S) = \mu(P,E(P),S)$.  

We will now prove the second part of the assertion. Let $P'$ be a
$\plc$ program and let $R$ and $S$ be nonempty and disjoint sets of
relation symbols appearing in $P'$. By Theorem \ref{datan}, it is
enough to show that the search problem $\mu(P',R,S)$ belongs to the 
class NP-search. This is, however, straightforward. A nondeterministic 
Turing machine $M$ (as defined in \cite{gj79}) for solving $\mu(P',R,S)$ 
can be described as follows:
\begin{enumerate}
\item Given an instance $D \in {\cal I}(R)$, $M$ grounds (in a 
deterministic way) the data-program pair $(D,P')$. Since $P'$ is 
fixed, the task can accomplished in polynomial time with respect to 
the size of $D$ (measured as the total number of symbols in $D$).
\item $M$ generates in a nondeterministic fashion (using its guessing 
module) a subset of the Herbrand base. This task involves the number of 
guesses that is not greater than $|\HB(D\cup P')|$, again a polynomial 
in the size of $D$. 
\item Next, $M$ checks (deterministically) that the subset that was 
guessed is a model of the ground theory. This task can be accomplished 
in time that is polynomial in the size of the grounding of the 
data-program pair $(D,P)$ which, as we already pointed out, is polynomial 
in the size of $D$.
\item If the subset that was guessed is not a model, $M$ moves to halting 
state NO. Otherwise, $M$ rewrites the contents of the tape so that only 
these ground atoms of the supported model that are built of relation 
symbols in $S$ are left, and moves to halting state YES.
\end{enumerate}
It is clear that tape contents for accepting computations are precisely 
projections of models of the data-program pair $(D,P)$ onto $S$. That
is, $M$ solves $\Pi$ nondeterministically in polynomial time. It
follows that $\mu(P',R,S)$ is in the class NP-search. 
\hfill $\Box$

\begin{corollary}
A search problem $\Pi$ is in the class NP-search if and only if there
is a finite $\plc$ program $P$ and nonempty disjoint sets $R$ and $S$
of relation symbols appearing in $P$ such that $\Pi=\mu(P,R,S)$.
\end{corollary}

Decision problems can be viewed as special search problems. Thus, in
particular, every decision problem in the class NP can be expressed by
means of a finite $\plc$ program (and two nonempty disjoint sets of 
relation symbols appearing in it). This observation is a counterpart 
to a result by Schlipf concerning $\datan$ \cite{sch91}.

\begin{corollary}
A decision problem $\Pi$ is in the class NP if and only if there
is a finite $\plc$ program $P$ and nonempty disjoint sets $R$ and $S$
of relation symbols appearing in $P$ such that $\Pi=\mu(P,R,S)$.
\end{corollary}

\section{Extensions of the logic $\plc$}
\label{eplc}

From the programming point of view, the logic $\plc$ provides a limited 
repertoire of modeling means: constraints must be represented as rules
(essentially, standard clauses of predicate logic). We will now present 
ways to enhance the effectiveness of logic $\plc$ as a programming 
formalism. Namely, we will introduce extensions to the basic formalism 
of the logic $\plc$ to provide direct support representations of some 
common ``higher-level'' constraints. We denote this extended logic
$\plc$ by $\plcp$.

\subsection{Adding cardinality atoms}

When considering the $\plc$ theories developed for the $n$-queens 
and vertex-cover problems one observes that these theories could be 
simplified if the language of the logic $\plc$ contained direct means 
to capture constraints such as: ``exactly one element is selected'' or 
``at most $k$ elements are selected''. 

We already noted in the introduction that extensions of the language 
of DATALOG$^\neg$ with explicit constructs to model such constraints
and the corresponding modifications in the algorithms to compute stable
models resulted in significant performance improvements. These gains
can be attributed to the fact that programs in the extended language
are usually much more concise, their ground versions use fewer variables
and have smaller sizes. Thus, the search space of candidate models
is also smaller.

It is natural to expect that similar gains are also possible in the
case of our formalism. With this motivation in mind, we extend the 
language of the logic $\plc$ by cardinality atoms. We we first 
consider a propositional language specified by a set of atoms $\at$. 
By a {\em propositional cardinality atom} (propositional c-atom, for
short), we mean any expression of the form $m\{p_1,\ldots,p_k\}n$ 
(one of $m$ and $n$, but not both, may be missing), where $m$ and $n$ 
are non-negative integers and $p_1 ,\ldots,p_k$ are atoms from $\at$. 
The notion of a rule generalizes in an obvious way to the case when 
propositional c-atoms are present in the language. Namely, a {\em 
c-rule} is an expression of the form
\[
C=\ \ A_1\wedge\ldots \wedge A_s\rra B_1\vee\ldots\vee B_t,
\]
where all $A_i$ and $B_i$ are (propositional) atoms or c-atoms.

Let $M\subseteq \at$ be a set of atoms. We say that $M$ {\em satisfies}
a c-atom $m\{p_1,\ldots,p_k\}n$ if
\[
m \leq |M\cap \{p_1,\ldots,p_k\}|\leq n.
\]
If $m$ is missing, we only require that $|M\cap \{p_1,\ldots,p_k\}|\leq
n$. Similarly, when $n$ is missing, we only require that $m \leq |M\cap
\{p_1,\ldots,p_k\}|$. A set of atoms $M$ {\em satisfies} a c-rule $C$
if $M$ satisfies at least one atom $B_j$ or does not satisfy at least
one atom $A_i$.

For example, if $\at=\{a,b,c,d\}$, then the expression
\[
a \rra 2\{a,c,d\}\vee d
\]
is a clause. The set $M=\{a,c\}$ is its model while $M'=\{a,b\}$ is
not.

To generalize the idea of a cardinality atom to the language of
predicate calculus, we need a syntax that will facilitate concise
representations of sets. To this end, we will adapt the standard
set-theoretic notation, where
\[
\{p(x)\colon x\in X\ \mbox{and}\ q(x)\}
\]
denotes the set of all atoms of the form $p(x)$ for which $x\in X$ and
$q(x)$ holds. For instance, in the language used for modeling the 
$n$-queens problem an expression
\[
\{q(R,C)\colon \indx(C)\ \mbox{and}\ R\leq C\}
\]
will be interpreted as a template for defining sets. For every ground
instantiation $r$ of its ``free'' variable $R$, it gives rise to the 
set ($n$ is a constant defined in the data file of the $n$-queens
data-program pair)
\[
\{q(r,c)\colon c=r, r+1,\ldots,n\}.
\]

Formally, we define a cardinality atom or {\em c-atom},
for short, to be any expression 
\[
l\{S_1;S_2;\ldots; S_k\}u,
\]
where 
$l$ and $u$ are terms (constants or variables) and 
$S_1, S_2, \ldots,S_k$ are {\em set definitions}. Intuitively, the 
meaning of a c-atom $l\{S_1;S_2;\ldots;S_k\}u$ is that at least $l$ 
and no more than $u$ of the atoms specified by set definitions $S_1,
\ldots,S_k$ are true. We will now make this intuition precise. Our
definitions are similar to those proposed in \cite{sn03} in the context
of SLP.

A {\em set definition} is an expression of the form
$p(t):d_1(s_1)\wedge \ldots\wedge d_m(s_m)$, where $p$ is a 
program relation symbol, $d_i$, $1\leq i\leq m$, are data or 
predefined relation symbols, and $t$, $s_i$, $1\leq i\leq m$, are 
tuples of terms. We note that it is possible that $m=0$. We also note 
that this concept is defined only in the context of data-program 
pairs as in that case there is a clear distinction between data and 
program predicates. A variable appearing in a set definition as an 
argument of one of data relation symbols is {\em bound}. Other 
variables appearing in this set definition are {\em free}. 

Let $S = p(t):d_1(s_1)\wedge \ldots\wedge d_m(s_m)$ be a set 
definition 
appearing in a data-program pair $T$. By our assumption, $T$ 
contains at least one constant. For every ground substitution 
$\vartheta$ whose domain contains all free variables in $S$ 
and does not contain any bound variables from $S$, by $S\vartheta$ 
we denote the set of atoms defined as follows: $S\vartheta$ is 
the set of all atoms of the form $p(t\vartheta\vartheta')$, where 
$\vartheta'$ is a ground substitution with the domain consisting 
of all bound variables in $S$ such that for every $i$, $1\leq 
i\leq m$, $d_i(s_i\vartheta\vartheta')$ holds (we recall that 
data and predefined relation symbols are fully specified by a 
data-program pair and this latter condition can be verified 
efficiently). We also note that if $m=0$, all variables appearing
in $t$ are free and $S\vartheta =\{p(t\vartheta)\}$.

Let us now consider a c-atom $A= l\{S_1;\ldots;S_k\}u$ appearing
in a theory or a data-program pair $T$. Without loss of generality,
we will assume that sets of bound and free variables appearing in 
c-atoms in $T$ are disjoint. 
Let $\vartheta$ be a ground substitution whose domain does not 
contain any bound variables appearing in $S_1,\ldots, S_k$. We define 
$A\vartheta$ as follows:
\begin{enumerate}
\item $A\vartheta = \bot$, if $l\vartheta$ or $u\vartheta$ are not
integers appearing as constants in $T$
\item $A\vartheta= l\vartheta\{S_1\vartheta\cup \ldots\cup 
S_k\vartheta\}u\vartheta$, otherwise. In this case, $l\vartheta$ 
and $u\vartheta$ are integer constants appearing in $T$,
and $S_1\vartheta\cup \ldots\cup S_k\vartheta$ is the set of ground 
atoms. 
\end{enumerate}
We define $\gr(T)$ as in Section \ref{plc}, with the stipulation that
c-atoms are grounded as specified above. The ground theory $\gr(T)$
consists of propositional c-rules. We define a set $M$ of ground
atoms to be a {\em model} of $T$ if it is a model of $\gr(T)$.

We will now illustrate these definitions with an example. Let $(D,P)$ 
be a data-program pair (in the language extended with c-atoms). Let us
assume that 
\begin{quote}
$D=\{d_1(1),d_1(2),d_1(3),d_2(a),d_2(b)\}$
\end{quote}
and that 
\begin{quote}
$C=\ \ d_1(X) \rra X\{p(X,Y)\colon d_1(Y)\wedge Y\geq X;\ \ q(Z)\colon
d_2(Z)\}$
\end{quote}
is a rule in $P$. The variables $Y$ and $Z$ are bound in $C$, the
variable $X$ is free.
Clearly, for every ground substitution $\vartheta$
such that $X\vartheta = a$ or $b$, both the antecedent and the 
consequent of the rule ground to $\bot$ and the rule grounds to
\begin{quote}
$\bot \rra \bot$.
\end{quote}
In every other ground substitution, $X$ is replaced with 1, 2 or 3.
Thus, we get the following three templates for propositional rules:
\begin{quote}
$d_1(1) \rra 1\{p(1,Y)\colon d_1(Y)\wedge Y\geq 1; q(Z)\colon
d_2(Z)\}$\\
$d_1(1) \rra 2\{p(2,Y)\colon d_1(Y)\wedge Y\geq 2; q(Z)\colon
d_2(Z)\}$\\
$d_1(1) \rra 3\{p(3,Y)\colon d_1(Y)\wedge Y\geq 3; q(Z)\colon
d_2(Z)\}$.
\end{quote}
Set definitions in each of these rules specify sets of ground atoms
and give rise to the following three ground instances of the rule 
$C$:
\begin{quote}
$d_1(1) \rra 1\{p(1,1), d_1(1,2), d_1(3), q(a), q(b)\}$\\
$d_1(2) \rra 2\{p(2,2), p(2,3), q(a), q(b)\}$\\
$d_1(3) \rra 3\{p(3,3), q(a), q(b)\}$.
\end{quote}
From the last of these rules it follows that the atoms $p(3,3)$,
$q(a)$ and $q(b)$ must be true in every model of $(D,P)$. 

In the extended logic $\plcp$ we can encode the vertex cover problem 
in a more straightforward and more concise way. Namely, there is no
the need for integers to represent indices as sets are represented
directly and not in terms of sequences! In this new representation 
$(D'_\vc(G,k),P'_\vc)$, $D'_\vc(G,k)$ is given by
\[
D'_\vc(G,k) = \{\vtx(v)\colon v\in V\} \cup \{\edge(v,w)\colon \{v,w\}
\in E\} \cup \{\size(k)\},
\]
and $P'_\vc$ consists of the clauses:

\setcounter{ct2}{0}
\begin{list}{VC$'$\arabic{ct2}:\ }{\usecounter{ct2}\topsep 0.03in
\parsep 0in\itemsep 0in}
\item $\invc(X)\rra \vtx(X)$
\item $\size(K) \rra \{\invc(X)\colon \vtx(X)\}K$ 
\item $\edge(X,Y) \rra \invc(X)\vee \invc(Y)$. 
\end{list}

Atoms $\invc(x)$ that are true in a model of the $\plc$ theory $(D'_\vc,
P'_\vc)$ define a set of vertices that is a candidate for a vertex cover.
(VC$'$2) guarantees that no more than $k$ vertices are included. 
(VC$'$3) enforces the vertex-cover constraint.

Cardinality atoms also yield alternative encodings to the graph-coloring
and $n$-queens problems. In both cases, we use the same representation of
input data and modify the program component only. In the case of the 
graph-coloring problem, a single rule, (C$'$3), directly stating that every 
vertex is assigned exactly one color, replaces two old rules (C3) and 
(C4).

\setcounter{ct2}{0}
\begin{list}{C$'$\arabic{ct2}:\ }{\usecounter{ct2}\topsep 0.03in
\parsep 0in\itemsep 0in}
\item $\clr(X,C)\rra \vtx(X)$
\item $\clr(X,C)\rra \col(C)$
\item $\vtx(X) \rra 1\{\clr(X,C)\colon \col(C)\}1$
\item $\edge(X,Y)\wedge \clr(X,C) \wedge \clr(Y,C) \rra \bot$.
\end{list}

In the case of the $n$-queens problem the change is similar. The rules
(nQ3) and (nQ4) are replaced with a single rule (nQ$'$3) and the rules
(nQ5) and (nQ6) with a single rule (nQ$'$4). 

\setcounter{ct17}{0}
\begin{list}{nQ$'$\arabic{ct17}:\ }{\usecounter{ct17}\topsep 0.03in
\parsep 0in\itemsep 0in}
\item $q(R,C) \rra \indx(R)$
\item $q(R,C) \rra \indx(C)$
\item $\indx(R) \rra 1\{q(R,C)\colon \indx(C)\}1$
\item $\indx(C) \rra 1\{q(R,C)\colon \indx(R)\}1$
\item $\indx(R) \rra \{q(R+I-1,I): \indx(I)\}1$
\item $\indx(C) \rra \{q(I,C+I-1): \indx(I)\}1$
\item $\indx(R) \rra \{q(R-I+1,I): \indx(I)\}1$
\item $\indx(C) \rra \{q(n-I+1,C+I-1): \indx(I)\}1$.
\end{list}

The rule (nQ$'$5) enforces the condition that the main ascending 
diagonal and all ascending diagonals above it contain at most one 
queen. The rule (nQ$'$6) enforces the same condition for the
ascending main diagonal and all ascending diagonals below it. Finally,
the rules (nQ$'$7) and (nQ$'$8) enforce the same condition for descending 
diagonals. In the original encoding we used only two clauses to represent 
these conditions. We could use them here again. However, the four
clauses that we propose here, and that are possible thanks to the
availability of c-atoms, result in significantly smaller ground
theories. We address this issue in detail in Section \ref{comp-sec}.

\subsection{Adding closure computation to logic $\plcp$}

In Section \ref{eps}, we presented programs capturing the concepts of 
reachability in graphs and of transitive closure of binary relations. 
These representations are less elegant and, more importantly, less 
concise than representations possible in SLP. For instance, the 
transitive closure of a binary relation $r$ can be computed by the 
following DATALOG$^\neg$ program:

\setcounter{ct2}{0}
\begin{list}{TC$'$\arabic{ct2}:\ }{\usecounter{ct2}\topsep 0.03in
\parsep 0in\itemsep 0in}
\item $tc(X,Y) \lla r(X,Y)$
\item $tc(X,Y) \lla r(X,Z), tc(Z,Y)$.
\end{list}
This encoding capitalizes on the minimality that is inherent in the 
stable-model semantics (in this case, the program, being a Horn 
program, has a unique {\em least} model). Moreover, the grounding of 
this program has size linear in the cardinality of the relation $r$.

Constraints involving reachability, transitive closure and other 
related concepts are quite common. In the problem of existence of a 
Hamiltonian cycle in a directed graph, we may first constrain candidate 
sets of edges to those that span collections of disjoint cycles 
covering all vertices in the graph (for instance, by imposing the 
restriction that in each vertex exactly one edge from the candidate 
set starts and exactly one edge from the candidate set ends). Clearly, 
such a candidate set is a Hamiltonian cycle if and only if it is 
connected. This requirement can be enforced by the constraint that all 
graph vertices be {\em reachable}, by edges in the candidate set, from 
some (arbitrary) vertex in the graph.  

With this motivation in mind, we will now introduce yet another 
extension of the basic logic providing, in particular, means to 
express constraints involving reachability, connectivity, transitive 
closure and similar related concepts in a way they are used in SLP. To this 
end, we extend both the syntax and the semantics of the logic $\plcp$. 

As it is standard, by a {\em Horn} rule we mean a $\plc$ rule (a rule 
without cardinality atoms) whose consequent is a single {\em regular} 
atom (that is, not an e-atom). {\em Horn} rules play a key role in this 
extension of the logic. The idea is to split the program component in a 
data-program pair into three parts. Intuitively, the first of them will 
describe initial constraints on the space of candidate solutions. The 
second, consisting of Horn rules, will ``close'' each candidate generated 
by the first part. The third component will provide additional constraints 
that have to be satisfied by the closure.

Formally, by an {\em extended program} we mean a triple $(G,H,V)$ such 
that
\begin{enumerate}
\item $G$ and $V$ are collections of (arbitrary) $\plcp$ rules, called 
{\em generating} and {\em verifying} rules, respectively, and $H$ is a 
collection of Horn rules 
\item No relation symbol appearing in the consequent of a rule in $H$ 
appears in rules from $G$.
\end{enumerate}
An extended data-program pair is a pair $(D,P)$, where $D$ is a set of 
ground atoms (data) and $P$ is an extended program. When listing an
extended program, we use the following convention. We write Horn rules
as in logic programming, starting at the left with the head, followed by 
the (reversed) arrow $\lla$ as the implication connective and, finally, 
followed by the conjunction of the atoms of the body. There is no need 
to explicitly distinguish between rules in $G$ and $V$ as the partition
is implicitly defined by $H$. Namely, non-Horn rules involving relation
symbols appearing in the consequents of Horn rules form the set $V$. All
other non-Horn rules form the set $G$.

Let $(D,P)$ be an extended data-program pair, where $P=(G,H,V)$. A set 
of ground atoms from the Herbrand base of $(D,G)$ is a {\em model} of 
$(D,P)$ if
\begin{enumerate}
\item $M$ is a model of $(D,G)$
\item the {\em closure} of $M$ under $H$, that is, the least Herbrand 
model of the Horn theory $M\cup H$, satisfies all ground instances of
rules in $V$.
\end{enumerate}
The first condition enforces that models of $(D,P)$ satisfy all 
constraints specified by $G$. Thus, $G$ can be regarded as a {\em 
generator} of the search space, as there are still additional constraints
to be satisfied. The second condition eliminates all these models 
generated by $G$ whose closure under $H$ violates some of the constraints 
given by $V$. In other words, $H$ computes the closure and $V$ {\em 
verifies} whether the closure has all of the desired properties.
 
As an illustration of the way this extension of logic $\plcp$ can 
be used we will provide a formal representation of the Hamiltonian-cycle 
problem, capturing intuitions described above. Let $G=(V,E)$ 
be a directed graph and let $v_0$ be an arbitrary vertex in $V$. To 
represent this data we set 
\[
D'_\hc(G,v_0) = \{\vtx(v)\colon v\in V\} \cup \{\edge(v,w)\colon \{v,w\}
\in E\} \cup \{\start(v_0)\}.
\]
Formally speaking, for the Hamiltonian-cycle problem, there is no
need to include $v_0$ in the data set. We do it, as our encoding 
involves the notion of reachability, for which some arbitrary 
``starting'' point is needed. The (extended) program part, $P'_\hc$,
consists of 
the following five rules. 
\setcounter{ct23}{0}
\begin{list}{HC$'$\arabic{ct23}:\ }{\usecounter{ct23}\topsep 0.03in
\parsep 0in\itemsep 0in}
\item $\hce(X,Y) \rra edge(X,Y)$
\item $1\{\hce(Y,X)\colon \vtx(Y)\}1$.
\item $1\{\hce(X,Y)\colon \vtx(Y)\}1$.
\item $\visit(Y) \lla \visit(X)\wedge \hce(X,Y)$
\item $\visit(Y) \lla \start(X)$. 
\item $\visit(X)$. 
\end{list}

We note the use of our notational convention. Clearly, the 
rules (HC$'$4) and 
(HC$'$5) form the Horn part (it is indicated by the way they are written). 
It follows that the rules (HC$'$1)-(HC$'$3) are generating and the rule 
(HC$'$6) is verifying. Intuitively, the rule (HC$'$1) guarantees that if an 
atom $\hce(x,y)$ is true then, $(x,y)$ is an edge (in other words, only 
edges of the graph can be chosen to form a Hamiltonian cycle). Rule 
(HC$'$2) captures the constraint that for every vertex $x$ there is 
exactly one selected edge that ends in $x$. Similarly, the rule (HC$'$3) 
captures the constraint that for every vertex $x$ there is exactly one 
selected edge that starts in $x$. Thus, every model of the data-program 
pair consisting of $D_\hc(G,v_0)$ and the rules (HC$'$1)-(HC$'$3) contains 
$D_\hc(G,v_0)$ and a set of atoms $\hce(x,y)$ that describe a particular 
selection 
of edges and that span in $G$ disjoint cycles covering all its 
vertices. Rules (HC$'$3) and (HC$'$4) define the relation $\visit$ that 
describes all vertices in $G$ reachable from 
$v_0$ by means of selected edges. Finally, the last rule verifies that 
all vertices are reached, that is, that selected edges form, in fact, 
a Hamiltonian cycle. 

\subsection{Expressive power of extended logics}

We close this section with an observation on the expressive power of
the logic $\eplc$. Since it is a generalization of the logic $\plc$,
it can capture all problems that are in the class NP-search. On the
other hand, the search problem of computing models of a data-program
pair $(D,P)$, where $P$ is a fixed $\eplc$ program, is an NP-search 
problem (a simple modification of the proof of the second assertion of 
Theorem \ref{ep-key} demonstrates that). Thus, it follows that the 
expressive power of the logics $\eplc$ does not extend beyond the class
NP-search. In other words, the logic $\eplc$ also captures the class 
NP-search. 

\section{Computing with $\plcp$ theories}
\label{comp-sec}

In the preceding sections, we focused on the use of the logic 
$\plcp$ as a language for encoding (programming) search problems 
and established its expressive power. In order to use the logic 
$\plcp$ as a computational problem solving tool we need algorithmic 
methods for processing data-program pairs and finding their models.

Let us recall that a set $M$ of ground atoms is a model of a 
data-program pair $(D,P)$ if and only if $M$ is a model of the theory 
$\gr(\cl(D)\cup P))$. Thus, to compute models one could proceed in two
steps: first, compute $\gr(\cl(D)\cup P))$ and then, find models of 
the ground theory. We refer to these steps as {\em grounding} and {\em 
solving}, respectively. This two-step approach is used successfully 
by all current implementations of SLP including, {\em smodels} and {\em 
dlv}. We will adhere to it, as well.

It is easy to see that the data complexity of grounding is in the class
P. That is, there is an algorithm that, for every data-program pair 
$(D,P)$ computes $\gr(\cl(D)\cup P))$ and, assuming that $P$ is fixed, 
works in time that is polynomial in the size of $D$. For instance, a 
straightforward enumeration of all substitutions of appropriate arities 
(determined by the numbers of free variables in program rules) can be 
adapted to yield a polynomial-time algorithm for grounding.

This straightforward approach can be improved. The size of grounding
(although polynomial in the size of the data part) is often very big.
To address this potential problem, we note that to compute the models
it is not necessary to use $\gr(\cl(D)\cup P))$. Any propositional 
theory that has the same models as $\gr(\cl(D)\cup P))$ can be used 
instead. In this context, let us note that the truth values 
of all ground atoms appearing in $\gr(\cl(D)\cup P))$ that are built of 
data relation symbols can be computed efficiently by testing whether 
they are present in $D$. Similarly, we can effectively evaluate truth 
values of all ground atoms built of predefined relation symbols by, 
depending on the relation, checking whether two constants are identical, 
different or, in the case of integer constants, whether one is the sum, 
product, etc. of two other integer constants.

Thus, the theory $\gr(\cl(D)\cup P))$ can be simplified by taking 
into account the truth values of ground atoms built of data and 
predefined relation symbols. Let $A$ be such a ground atom. 
\begin{enumerate}
\item If $A$ appears in the consequent of the clause and is true, 
we eliminate this clause
\item If $A$ appears in the consequent of the clause and is false, 
we eliminate $A$ from the consequent of the clause
\item If $A$ appears in the body of a clause and is true,
we eliminate $A$ from the body.
\item If $A$ appears in the body of the clause and is false, we 
eliminate this clause.
\end{enumerate}

These simplifications may reveal other atoms with forced truth values
and the process continues, much in the spirit of unit propagation used 
in satisfiability solvers. For instance, if we obtain a rule consisting 
of a single (regular) atom and this atom appears in the consequent of 
the rule, the atom must be true. If, on the other hand, this single 
atom appears in the body of the rule, it must be false. Furthermore, if 
a cardinality atom of the form $m\{p_1,\ldots,p_k\}$n, 
is forced to be true and the number of atoms $p_i$ that have been 
already assigned value true is $m$, then all the unassigned atoms $p_i$
must be false. In addition, if the number of atoms $p_i$ that have been
already assigned value false is $k-m$, then all unassigned atoms must 
be true. Similar propagation rules exist for the case when a c-atom is
forced to be false.

We continue the process of simplifying the
theory as long as new atoms with forced truth values are discovered.
We call the theory that results when no more simplifications are possible
the {\em ground core} of a data-program pair $(D,P)$. We denote it by
$\core(D,P)$. 

We have the following straightforward result (as in the other cases 
before, we do not explicitly mention ground predefined atoms when 
specifying models).

\begin{proposition}
\label{core}
Let $(D,P)$ be a data-program pair. A set $M$ of ground atom is a model
of $(D,P)$ if and only if $M= D\cup T\cup M'$, where $T$ is the set of 
atoms that are forced to be true and $M'$ is a model of $\core(D,P)$.
\end{proposition}

Proposition \ref{core} suggests that for the grounding step, it is 
enough to compute $\core(D,P)$ rather than $\gr(\cl(D)\cup P))$. It is
an important observation. The size of the theory $\core(D,P)$, measured
as the total number of symbol occurrences, is usually much smaller when 
compared to that of $\gr(\cl(D)\cup P))$. Following this general idea, 
we designed and implemented a program, {\em psgrnd} that, given a 
data-program pair $(D,P)$, computes its ground equivalent $\core(D,P)$. 

We will now focus on the second step --- searching for models of a
propositional $\plcp$ theory. First, we will consider the class of 
theories that are obtained by grounding data-program pairs whose 
program component does not contain c-atoms. In this case, the ground 
core of a data-program pair is a collection of standard propositional 
clauses (written as implications). The program {\em psgrnd} provides
an option that, in such case, produces the ground core of the input 
data-program pair in the DIMACS format. Consequently, most of the 
current implementations of propositional satisfiability (SAT) solvers 
can be used in the solving step to compute models. Thus, we can view 
the logic $\plc$ as a programming tool for modeling 
problems in terms of propositional constraints and regard {\em psgrnd} 
as a front-end facilitating the use of SAT solvers.

If c-atoms and Horn rules are present in a program, the theory after 
grounding and simplification is a propositional $\plcp$ theory that 
contains, in general, (propositional) c-atoms and propositional Horn
rules. Thus, SAT solvers are not directly applicable. One approach in 
such case is to represent c-atoms and closure rules by means of 
equivalent (standard) propositional theories. It is possible since, 
as we noted earlier, logics $\plc$ and $\plcp$ have the same expressive 
power.

We argue, however, that a more promising approach to compute models 
of data-program pairs is to design solvers for propositional $\plcp$ 
theories that are direct outcomes of the grounding process and, in 
general, may contain c-atoms. The reason is that using high-level 
constraints results in programs whose ground representations are often 
more concise then those obtained by corresponding programs that do not 
involve such constraints. We will illustrate this point using 
programs developed earlier in the paper.

We start with the vertex-cover problem. Let $G$ be an input graph with
$n$ vertices and $m$ edges, and let $k$ be an integer $k$ specifying
the cardinality of a vertex cover. In the case of the program
consisting of rules (VC1) - (VC5), our grounding algorithm results in 
a propositional theory with $kn$ atoms of the form $\vc(i,x)$ and with 
$\Theta(kn^2)$ rules of total size (measured by the number of atom 
occurrences) also $\Theta(kn^2)$. On the other hand, grounding of the 
program consisting of rules (VC$'$1) - (VC$'$3) yields a theory with 
$n$ atoms of the form $\invc(x)$ and with $\Theta(m)$ rules of total 
size $\Theta(n+m)$. Thus, this latter encoding involves fewer atoms (if 
$k\geq 2$) and has the size that is asymptotically smaller.

Next, we will consider the Hamiltonian-cycle problem. Our first
encoding (rules (HC1) - (HC6)) grounds to a theory with $n^2$ 
atoms and with total size $\Theta(n^2 + n(n^2-m))$. Our second 
encoding, involving Horn rules (rules (HC$'$1) - (HC$'$7)), grounds 
to a theory with $n^2 +n$ atoms and the total size of $\Theta(n^2)$. 
Thus, even though this theory uses slightly more atoms, it has 
significantly smaller total size (except for the case of ``almost 
complete'' graphs, that is, graphs with the number of ``missing''
edges equal to $o(n^2)$, one-order of magnitude smaller).

For the original encoding of the $n$-queens program, {\em psgrnd}
produces a propositional theory of size $\Theta(n^3)$. On the other
hand, it is easy to see that grounding of the the second encoding
(the one involving cardinality atoms), has size $\Theta(n^2)$ --- a
gain of an order of magnitude.

In the case of the encodings for the graph-coloring problems, we 
also obtain more concise theories by grounding programs designed 
with the use of c-atoms. Indeed, the rule (C$'$3) grounds to a 
smaller theory than rules (C3) and (C4). The improvement is, in 
general, by a constant factor and so, it is not asymptotically 
better.

Since encodings involving c-atoms are usually smaller and define 
smaller search spaces, it is important to design solvers that can take
direct advantage of these small representations. We developed a solver, 
{\em aspps} (short for ``answer-set programming with propositional 
schemata''), that can directly handle c-atoms and closure rules. 
The {\em aspps} solver is an adaptation of the Davis-Putnam algorithm 
for computing models of propositional CNF theories. That is, it is a
backtracking search algorithm whose two key components are {\em unit 
propagation} and {\em branching}. Unit propagation ``propagates'' 
through the theory truth values established so far. If there is a rule 
with all atoms in the antecedent assigned value true and all but one 
atom in the consequent assigned value false, then the remaining 
``unassigned'' atom in the consequent must be true for the rule to 
hold. Similarly, if all atoms in the consequent of a rule are false 
and if all but one atom in the antecedent are true, the only 
``unassigned'' atom in the antecedent must be false. In this way any 
partial assignment of truth values to atoms {\em forces} truth 
assignments on some additional atoms. When no more atoms can be forced,
the second module, {\em branching}, selects a way to split search space
into separate parts. When search in a part fails, the program backtracks 
and tries another. 
   
A key difference between {\em aspps} and satisfiability algorithms is
in how {\em branching} is implemented. In satisfiability solvers, 
in order to branch, we pick an atom, say $a$, and split the search space 
into two parts. In one of them we assume that the atom $a$ is true. In 
the other one we assume that $a$ is false. Propositional $\plcp$ 
theories may, in general, contain c-atoms and {\em aspps} considers
them too when selecting a way to branch. 

To explain the method that {\em aspps} uses, let us observe that the 
unit propagation may, in particular, assign a truth value to a c-atom 
appearing in the theory. That constrains possible truth assignments to 
unassigned atoms that form the c-atom. 

For example, let us consider a propositional c-atom $C= 1\ \{ a, b, 
c , d\}\ 1$ that we know must be true. Let us assume that $d$ has 
already been assigned value false and that $a$, $b$ and $c$ have not.
There are exactly three ways in which atoms $a$, $b$ 
and $c$ can be assigned truth values consistent with $C$ being true:
\begin{quote}
   $a = \Tr, b = \Fa, c=\Fa$\\
   $a = \Fa, b = \Tr, c=\Fa$\\
   $a = \Fa, b = \Fa, c=\Tr$.
\end{quote}

It follows that if a truth value of a c-atom $C$ has been forced, we 
have an additional way to split the search. Namely, we can consider in 
turn each truth assignment to unassigned atoms appearing in $C$ that 
is consistent with the truth value of $C$. In our example, if $C$ is 
true, we could split the search space into three subspaces by assuming 
first that $a = \Tr$, $b = \Fa$ and $ c=\Fa$, then that $a = 
\Fa$, $b = \Tr$ and $c=\Fa$ and, finally, that $a = \Fa$, $b = \Fa$ and 
$c=\Tr$. 

The choice of the way to branch is of vital importance. To make this 
selection, the {\em aspps} program approximates the degree to which 
the atom is constrained. That is, {\em aspps} first assigns to each 
clause a weight based on its current length. The fewer atoms in a clause 
the more constraining it is and the greater its weight. The weight of a 
(regular) atom is defined as the sum of the weights of all clauses 
containing it. It is this number that {\em aspps} uses to estimate how 
much the atom is constrained.
 
When looking for the way to branch, the {\em aspps} program considers all 
(regular) atoms that have not been assigned a truth value yet. It also 
considers some c-atoms. Let $C$ be a c-atom that has been forced to be 
true by earlier choices. Let $A$ be the set of atoms appearing in $C$ 
that have not received a truth value yet. The atom $C$ is considered 
as a candidate to define branching if the number of truth assignments 
to atoms in $A$ that are consistent with the truth value of $C$ (the 
number of branches $C$ defines) is less than or equal to $|A|$.

If there are c-atoms satisfying these conditions, {\em aspps} will select
this one among them that maximizes the sum of weights of unassigned atoms 
that appear in it. Otherwise, {\em aspps} will branch on a regular atom 
with the maximum weight. If a propositional $\plcp$ theory contains Horn 
clauses, they play no role in the process of selection the next atom for 
branching.  They participate though in the unit propagation step.

The source codes, information on the implementation details
for programs {\em psgrnd} and {\em aspps} and on their use is 
available at \url{http://www.cs.uky,edu/ai/aspps/}.

\section{Experimental results}

Several data-program pairs that we presented in the paper (both
with and without c-atoms and Horn rules) show that the logics 
$\plc$ and $\plcp$ are effective as formalisms for modeling search 
problems. In this section we will demonstrate computational 
feasibility of these formalisms when combined either with our native 
solver {\em aspps}, specifically tailored to handle the syntax
of the logic $\plcp$ (c-atoms and Horn rules), or with off-the-shelf 
SAT solvers. 

We show that {\em aspps} is generally comparable in performance with 
that of {\em smodels} and, in the cases discussed here, even faster. 
We chose {\em smodels} for the comparison since (1) {\em smodels} 
accepts a similar syntax as {\em aspps}, (2) in the case of each 
problem considered here, there is an {\em smodels} program essentially 
identical in its basic structure to the $\plcp$ program presented 
in the paper, and (3) {\em smodels} is at present one of the most 
advanced implementations of answer-set programming paradigm
based on $\datan$ with the stable-model semantics. 

Next, we show
that our language of data-program pairs in the basic logic $\plc$ 
(without c-atoms and Horn rules), together with the program {\em 
psgrnd}, greatly simplifies the use of SAT solvers in computing 
solutions to search problems. 

Finally, we compare the performance 
of {\em aspps} and {\em smodels} with that of SAT solvers as engines
for solving search problems. 

We stress that our experiments did not aim at demonstrating 
superiority of one solver over another. That would require a much 
more comprehensive and careful experimental study. Our objective was
to demonstrate the feasibility of our approach.

For our test cases we selected problems that we used as examples
throughout the paper: the $n$-queens, graph-coloring, vertex-cover, 
and Hamiltonian cycle problem. We used {\em psgrnd/aspps} and {\em 
smodels} for encodings in the logic $\plcp$ (that is, encodings 
involving c-atoms and Horn rules). We used the combinations 
{\em psgrnd/satz} and {\em psgrnd/chaff} for programs without 
c-atoms and Horn rules. In the experiments we used the following
versions of these programs: {\em zchaff} \cite{chaff}, {\em satz215.2} 
\cite{satz}, {\em lparse-1.0.6} and {\em smodels-2.26} both at 
\cite{smodels}, {\em aspps.2001.10.18}, {\em psgrnd.2001.10.18} and, 
finally, {\em psgrnd.2002.10.11} (as a front-end for satisfiability 
solvers) \cite{aspps}. All our experiments were performed on a Pentium 
IV 1.7 GHz machine running linux. 

In the case of vertex cover, for each $n=50, 60, 70, 80$ we randomly 
generated 100 graphs with $n$ vertices and $2n$ edges. For each graph 
$G$, we computed the minimum size $k_G$ for which the vertex cover can 
be found. We then tested {\em aspps}, {\em smodels} and {\em satz} on 
all the instances $(G,k_G)$. The results (Table \ref{tab1}) represent the 
average execution times. Encodings we used for testing {\em aspps} and 
{\em smodels} where based on the program (VC$'$1) - (VC$'$3). For 
satisfiability solvers we used encodings based on the clauses (VC1) - 
(VC5) (as cardinality constraints cannot be handled by satisfiability 
solvers). 

As we observed, the size of the encoding (VC$'$1) - (VC$'$3) is, in
general, asymptotically smaller than that of (VC1) - (VC5). Thus,
satisfiability solvers had to deal with much larger theories (hundreds
of thousands of clauses for graphs with 80 vertices as opposed to 
a few hundred
when c-atoms are used). Consequently, they did not perform well.

As concerns {\em smodels} versus {\em aspps}, in general {\em aspps} is 
somewhat (about three times) faster than smodels and the difference seems
to grow with size.  

\begin{table}[h]
\begin{center}
{\small
\begin{tabular}{|l|r|r|r|r|}
\hline
$n$ & 50 & 60 & 70 & 80 \\
\hline
{\em aspps} & 0.011 & 0.070 & 0.463 & 1.996 \\
\hline
{\em smodels} & 0.043 & 0.116 & 1.584 & 8.157 \\
\hline
\end{tabular}
\caption{\small Timing results (in seconds) for the vertex-cover 
problem. Average 
time for each set of 100 random generated graphs; {\em satz} and {\em 
zchaff} were halted after 10 minutes on a single instance.}
\label{tab1}
}
\end{center}
\end{table}
     
For the $n$-queens problem, our solver performed exceptionally well.
Cardinality constraints play here again a crucial rule. {\em Aspps}
has to work with theories obtained by grounding the program (nQ$'$1) -
(nQ$'$8). In contrast, {\em satz} and {zchaff} have to work with much 
larger theories obtained by grounding the program (nQ1) - (nQ7). When 
$n=70$ this means the difference between 416 and 562030 rules,
respectively (in each case, the number of ground atoms is 4900).
Both solvers required therefore much more time than {\em
aspps}. In fact, we stopped {\em satz} after 10 minutes on the 40-queen 
instance. {\em Zchaff} performed much better than {\em satz} and 
completed the computation in the case of $n=70$ in just under 10 
minutes. Given the size of the theory it has to work with, this is 
quite remarkable. One lesson, we believe, is that there is still a 
large potential for improvements in the way {\em aspps} implements 
search. We also note that {\em aspps} and {\em zchaff} solvers both 
exhibited a somewhat irregular performance growth pattern as the number 
of queens increased. Lastly, we note that {\em smodels} did not 
complete computation for $n=40$ in the 10 minutes we allocated, despite 
the fact that we used a more concise encoding, similar to that 
processed by {\em aspps}. Table \ref{tab2} summarizes these results.

\begin{table}[h]
\begin{center}
{\small
\begin{tabular}{|l|r|r|r|r|r|}
\hline
\# of queens & 40 & 50 & 60 & 70 & 80  \\
\hline
{\em aspps} & 0.20 & 0.06 & 2.19 &0.31  &0.17   \\
\hline
{\em zchaff} & 158.92 & 157.74 & 283.87 & 558.10 &*** \\
\hline
\end{tabular}
}
\caption{\small Timing results (in seconds) for the $n$-queen problem; 
{\em satz} 
and {\em smodels} were halted after 10 minutes on the instance with
$n=40$.}
\label{tab2}
\end{center}
\end{table}

In the case of the graph colorability problem, as we observed in the
previous section, c-atoms do not give rise to significant gains in the 
size of the ground theory. Given the amount of research devoted to 
satisfiability solvers and still relatively few efforts to develop 
fast solvers for logics involving cardinality constraints, it is not 
surprising that satisfiability solvers outperform both {aspps} 
and {\em smodels}. Our results also show {\em satz} outperforming
{\em zchaff}, which may be attributed to the fact that our test graphs
were randomly generated and did not have any significant internal
structure that could be capitalized on by {\em zchaff}. As concerns 
{\em aspps} and {\em smodels}, they show essentially the same 
performance. We summarize the relevant results in Table \ref{tab3}.
The graphs for the 3-colorability problem were generated randomly 
with vertex/edge ratios such that approximately 1/2 of the graphs 
were 3-colorable. For each value $n=100$, 200 and 300, we generated 
a set of 1000 graphs. The values that we report are the average execution
times. 

\begin{table}[h]
\begin{center}
{\small
\begin{tabular}{|l||r|r|r|}
\hline
$n$ & 100& 200 & 300  \\
\hline
{\em aspps}& 0.006 &0.302 &  13.678 \\
\hline
{\em smodels}& 0.026 & 0.495 & 16.043  \\
\hline
{\em satz}&0.013 & 0.077 & 1.416 \\
\hline
{\em zchaff}&0.002 & 0.107 & 7.952 \\
\hline 
\end{tabular}
}
\caption{\small Timing results (in seconds) for the graph 3-coloring 
problem.}
\label{tab3}
\end{center}
\end{table}

Our last experiment concerned the problem of computing Hamiltonian
cycles. As in other cases when additional constraints (transitive
closure computation, in this case) result in much smaller theories,
both {\em smodels} and {\em aspps} outperform {\em satz} and {\em
zchaff}. In addition, in this case, {\em aspps} significantly 
outperforms {\em smodels}. In the experiments, we considered graphs 
with 20, 40, 60, 80 and 100 vertices with the number of edges chosen 
so that the likelihood of the existence of a Hamiltonian cycle is 
close to 0.5. For each set of parameters, we generated 1000 instances. 
The times given in Table \ref{tab4} represent average execution times.

\begin{table}[h]
\begin{center}
{\small
\begin{tabular}{|l||r|r|r|r|r|}
\hline
V/E & 20/75 & 40/180 & 60/300 & 80/425 & 100/550  \\
\hline
{\em aspps} & 0.000 & 0.001 & 0.002 & 0.003 & 0.005  \\
\hline
{\em smodels} & 0.006 & 0.034 & 0.117 & 0.255 & 0.456   \\
\hline
{\em satz} & 0.122 & *** & *** & *** & ***  \\
\hline
{\em zchaff} & 0.144 & 2.277 &33.405 & ***& ***  \\
\hline
\end{tabular}
}
\caption{\small Timing results (in seconds) for the determining presence of 
a Hamilton cycle in a graph.}
\label{tab4}
\end{center}
\end{table}

It is clear from these results that our solver {\em aspps} is
competitive with {\em smodels} and SAT solvers such as {\em zchaff} 
and {\em satz} as a processing back-end for problems encoded as
data-program pairs in the logics $\plc$ and $\plcp$.

\section{Conclusions}

Our work demonstrates that predicate logic and its extensions 
can support answer-set programming systems in a way in which stable 
logic programming does. To put it differently, we show that predicate
logic can be an effective declarative programming formalism. 

In the paper we described logic $\plc$ that can be used to uniformly 
encode search problems. We proved that the expressive power of this 
logic is given by the class NP-search. Thus, it is the same as the 
expressive power of $\datan$, even though it is conceptually
simpler --- its semantics is essentially that of propositional logic.

We demonstrated the use of our logic in modeling such search problems
as graph coloring, vertex cover, $n$-queens, Hamiltonian cycle and 
transitive closure. 

We designed a program {\em psgrnd}, that given a data-program pair 
in the logic $\plc$, encoding a search problem, computes its 
equivalent propositional representation in the DIMACS form. In this 
way, it becomes possible to compute models of data-program pairs and,
consequently, solve the corresponding search problems by means of
standard off-the-shelf satisfiability solvers. We demonstrated that the
approach is feasible and effective by applying {\em satz} and {\em 
zchaff} to propositional theories produced by {\em psgrnd}. 

We argued that the logic $\plc$ can benefit from extensions allowing
explicit representations of some commonly used constraint such as
cardinality constraints and transitive closure. Encodings of search
problems that take advantage of these extensions are usually much
smaller and, consequently, could result in smaller search spaces if
solvers capable to take advantage of direct representations of
high-level constraints were available. We designed one such solver,
{\em aspps}. Our experimental results are encouraging. {\em Aspps}
is competitive with {\em smodels}, a state-of-the-art processing 
engine for $\datan$ programs extended by cardinality constraints and 
other constructs. In fact, in several cases, {\em aspps} outperforms
{\em smodels}.

The results of the paper show that programming front-ends for 
constraint satisfaction problems that support explicit coding of complex 
constraints facilitate modeling and result in concise representations.
They also show that solvers such as {\em aspps} that take advantage of 
those concise encodings and process high-level constraints directly, 
without compiling them to simpler representations, exhibit very good 
computational performance. These two aspects are important. 
Satisfiability checkers often cannot effectively solve problems simply 
due to the fact that encodings they have to work with are large. For
instance, for the vertex-cover problem for graphs with 80 vertices and
160 edges, {\em aspps} has to deal with theories that consist of a few
hundred of rules only. In the same time pure propositional encodings 
of the same problem contain over one million clauses --- a factor that
undoubtedly is behind much poorer performance of {\em satz} and {\em
zchaff} on this problem.  
 
Our work raises new questions. Further extensions of logic $\eplc$ are 
possible. For instance, constraints that impose other conditions on 
set cardinalities than those considered here (such as, the {\em parity} 
constraint) might be included. We will pursue this direction. Similarly, 
there is much room for improvement in the area of solvers for the
propositional logic $\plcp$ and {\em aspps} can certainly be improved.
There is also a potential for developing local search techniques for 
the logic $\plcp$. The task seems much easier than in the case of 
$\datan$ programs, where finding successful local search algorithms 
turned out to be hard \cite{ds02}.

\section*{Acknowledgments}

This material is based upon work supported by the National Science
Foundation under Grants No. 9874764 and 0097278.

{\small


\newcommand{\etalchar}[1]{$^{#1}$}

}
\end{document}